\documentclass{ws-rv9x6}
\usepackage{subfigure}
\usepackage{ws-rv-thm}
\usepackage{slashed}
\usepackage[square]{ws-rv-van}
\makeindex

\newcommand{\I}{\mathrm{i}}
\newcommand{\Tr}{\text{Tr}}

\graphicspath{
{./}
{./figures/}
}

\begin{document}

\chapter[Spectral functions and in-medium properties of hadrons]{Spectral functions and in-medium properties of hadrons}
\label{JW_chapter}

\author{R.-A.~Tripolt}

\address{European Centre for Theoretical Studies in Nuclear Physics and related Areas (ECT*), Trento, Italy}

\author{L.~von Smekal}

\address{Justus-Liebig-Universit{\"a}t Gie{\ss}en, Germany}

\author[R.-A.~Tripolt, L.~von Smekal \& J.~Wambach]{J.~Wambach}

\address{European Centre for Theoretical Studies in Nuclear Physics and related Areas (ECT*), Trento, Italy, and\\
	Technische Universit{\"a}t Darmstadt, Germany}

\begin{abstract} 
The in-medium modifications of hadron properties such as masses and decay widths have been a major focus of the scientific work of Gerry Brown and the insights gained by him and his collaborators made them major drivers of this field for several decades. Their prediction of experimental signals in di-lepton pair production in relativistic heavy-ion collisions were instrumental in initiating large experimental campaigns which continue until today. In this chapter we review recent results which elucidate the relation of hadronic spectral properties at finite temperature and density to the restoration of spontaneously broken chiral symmetry.  
\end{abstract}

\body

\section{Introduction}
\label{sec:intro}

The spectral properties of hadrons for matter under extreme conditions in temperature and density (QCD matter) as encountered in the early universe and the core of neutron stars are of fundamental importance for the understanding of the relevant degrees of freedom in the equation of state and the transport properties. It was realized early on by Gerry and his collaborators that spontaneously broken chiral symmetry and its restoration play an important role in this context. Spontaneously broken chiral symmetry endows the up and down quarks with almost all of their mass as confirmed by QCD lattice studies of the quark propagator as well as by Dyson-Schwinger results \cite{Parappilly:2005ei,Williams:2015cvx}. It seems therefore rather natural to conclude that all hadrons composed of light quarks (except for the pion as a Goldstone boson) essentially loose most of their mass as chiral symmetry gets gradually restored. This has led to the conjecture of a scaling law of hadron masses with the in-medium chiral condensate \cite{Brown1991,Brown2004}. The discussions are far from over and they are at the very heart of the question of mass generation in QCD. In this context chiral effective quark Lagrangians have played an important role, as they are able to describe the quark mass generation through the chiral condensate. At the same time, hadrons (mesons and baryons) emerge as dressed quark-antiquark or three-quark composites. 

In the vacuum-spectra of hadrons containing light quarks, broken chiral symmetry manifests itself through the absence of doublets of opposite parity. As chiral symmetry gets restored with increasing temperature and (net) quark density the parity partners approach degeneracy and their spectral properties eventually become identical. For vector and axial-vector mesons this is encoded in the Weinberg sum rules \cite{Weinberg:1967kj, Kapusta:1993hq} for instance. How this degeneracy happens and whether it is accompanied by significant drops in (pole) masses is the subject of the present study. It will be found that this is a rather subtle issue, which we will comment on further in the end. We employ the Quark-Meson (QM) model (a variant of the Gell-Mann-Levy linear sigma model), which shares the most important chiral aspects with QCD. The aim is to compute the chiral restoration transition and the resulting medium modification of pions and their chiral partner (the sigma meson) as a function of temperature $T$ and quark chemical potential $\mu$ in a thermodynamically consistent manner. 

As the chiral transition may become a true phase transition for specific values of $T$ and $\mu$, a method is needed that is able to deal with phase transitions in a reliable manner. Such a method is the Functional Renormalization Group (FRG) which is widely used in quantum field theory and statistical physics \cite{Berges:2000ew,Polonyi:2001se,Pawlowski:2005xe,Schaefer:2006sr,Kopietz2010,Braun:2011pp, Friman:2011zz, Gies2012}. It goes beyond mean-field theory (MFT) by incorporating thermal and quantum fluctuations through scale-dependent evolution equations for the effective action of a given theory (in fact MFT is the starting point at the ultraviolet scale from which the evolution is initiated). In MFT, the thermodynamically consistent excitation spectrum is obtained from the `Random-Phase-Approximation (RPA)' (an infinite sum of one-loop diagrams) first introduced for the electron gas by D.~Bohm and D.~Pines in 1953 \cite{Bohm:1953zza}. Gerry together with M.~Bolsterli has made extensive use of the RPA in the context of nuclear collective excitations with devising an analytic model (the `schematic model' \cite{Brown:1959zzb}) which has provided valuable insight into the role of residual nuclear interactions in the collectivity of giant resonances and low-lying vibrational modes of spherical nuclei.

Although MFT often captures the gross features of the equilibrium properties of a given system, quantitative predictions and correct descriptions of phase transitions and critical phenomena require the proper inclusion of thermal and quantum fluctuations, in particular those of the order parameters due to collective excitations. However, a thermodynamically consistent calculation of spectral properties beyond MFT poses an inherently difficult problem as it involves real-time correlation functions.  Within the FRG, the thermal equilibrium state is usually obtained in an Euclidean approach, making use of the Matsubara formalism. To compute correlation functions in Minkowski space-time an analytic continuation is required. The analytic continuation problem is common to all Euclidean approaches to quantum field theory and describes the need to analytically continue from imaginary to real energies. This technical difficulty, for example, also arises in Lattice QCD where one has to use numerical methods to reconstruct real-time correlation functions from the Euclidean data, cf.~\cite{Vidberg:1977,Jarrell:1996,Asakawa:2000tr,Dudal2013}. The method  \cite{Strodthoff:2011tz, Kamikado2013, Tripolt2014, Tripolt2014a} applied in the following avoids this exponentially hard inverse problem as it deals with the analytic continuation on the level of the FRG flow equations which can then be solved for real frequencies without the need for any numerical reconstruction. 
 \section{The Functional Renormalization Group}
\label{sec:FRG}

The FRG represents a powerful non-perturbative continuum framework to study quantum field theories as statistical systems near criticality in Euclidean space-time. The basic idea is to implement Wilson's coarse-graining procedure \cite{Wilson1971, Wilson1974} within a functional formulation. One usually starts with an Ansatz for the classical or mean-field action $S$ to describe the microphysics at some high energy-momentum scale and then successively lowers this scale to include fluctuations with lower and lower momenta in order to arrive at the full quantum effective action $\Gamma$. The connection between the ultraviolet (UV) and the infrared (IR) regimes is established by an FRG flow equation for a scale-dependent effective action, the so-called effective average action.

We briefly recall that the effective action $\Gamma[\phi]$ is the generating functional for the one-particle-irreducible (1PI) correlation functions of a theory. It is obtained from the generating functional $Z[J]$ for the full Green functions which we formally write as an Euclidean path integral over some generic field~$\varphi$,
\begin{equation}
Z[J]=\int \mathcal{D}\varphi \:\exp\left( -S[\varphi]+\int d^4x\: J(x)\varphi (x)\right),
\end{equation}
where $S[\varphi]$ is the classical action. The generating functional for the connected $n$-point Green functions is then given by
\begin{equation}
W[J]=\log Z[J],
\end{equation}
and the effective action $\Gamma[\phi]$ is obtained as the Legendre transform of $W[J]$ with respect to the expectation value $\phi(x) = \langle \varphi(x) \rangle_J $ of the field  $\varphi(x)$ in presence of the source $J(x)$,
\begin{equation}
\Gamma[\phi]=\sup_J \left( \int d^4x \:J(x) \phi(x)-W[J]\right).
\end{equation}
When evaluated for a homogeneous equilibrium state with (spatially) constant $\phi= \phi_0$ which satisfies
\begin{equation}
\left.\frac{\delta\Gamma[\phi]}{\delta\phi}\right|_{\phi=\phi_0}=0,
\end{equation}
the effective action is related to the Grand Canonical potential by
\begin{equation}
\label{EffTP}
\Omega(T,\mu)=\frac{T}{V}\Gamma[\phi_0].
\end{equation}

\begin{figure}
	\centerline{{\includegraphics[width=0.32\textwidth]{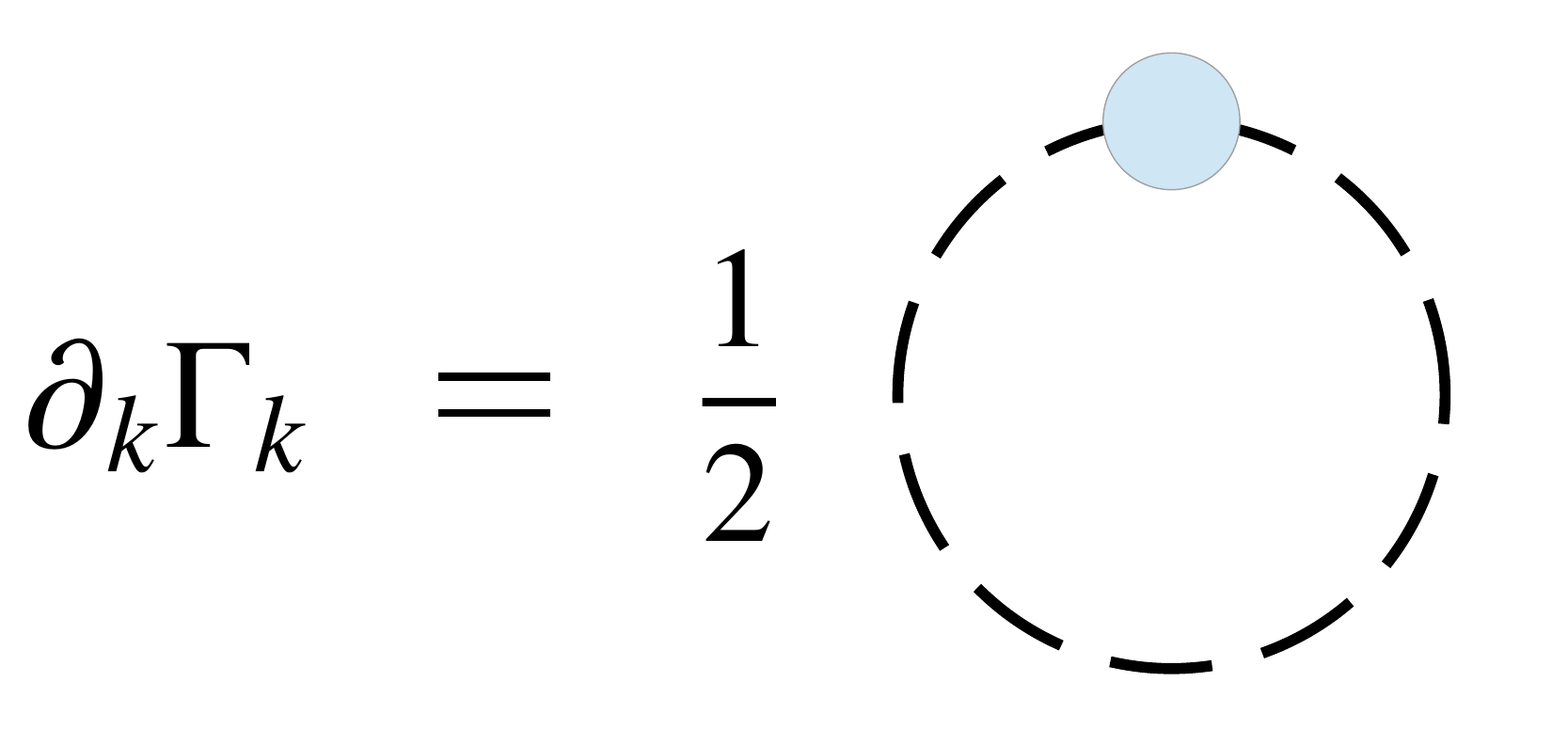}}}
	\caption{Diagrammatic representation of the flow equation for the effective average action, Eq.~(\ref{eq:Wetterich}). The dashed line represents the full field and scale-dependent propagator and the circle the regulator insertion $\partial_kR_k$.}
	\label{fig:flow_gamma_simple}
\end{figure}

Wilson's coarse-graining idea is implemented by introducing a regulator function $R_k$ which suppresses fluctuations of modes with momenta lower than the RG scale $k$. Successively lowering this scale from the UV to the IR, thermal and quantum fluctuations are taken into account momentum shell by momentum shell. The scale-dependent analogues of the generating functionals $W[J]$ and $\Gamma[\phi]$ are then given by
\begin{align}
W_k[J]&=\log \int \mathcal{D}\varphi\: \exp \left( -S[\varphi] -\Delta S_k[\varphi]+\int d^4x\: J(x)\varphi(x)\right),\\
\Gamma_k[\phi]&=\sup_J\left( \int d^4x\: J(x)\phi(x)-W_k[J]\right) -\Delta S_k[\phi],
\end{align}
where the regulator insertion $\Delta S_k[\phi]$ is given by
\begin{equation}
\Delta S_k[\phi]=\frac{1}{2}\int\frac{d^4q}{(2\pi)^4}\:\phi(-q)R_k(q)\phi(q),
\end{equation}
which acts as a scale-dependent mass term.
The effective average action $\Gamma_k$ interpolates between the classical action $S$ at the UV scale $\Lambda$ and the effective action $\Gamma[\phi]$ at $k=0$, i.e.
\begin{equation}
\Gamma_{\Lambda}\approx S,\qquad \lim_{k\rightarrow 0} \Gamma_k=\Gamma,
\end{equation}
The scale dependence of $\Gamma_k$ is given by the following exact flow equation, also known as the `Wetterich equation' \cite{Wetterich:1992yh, Morris1994}, 
\begin{equation}
\label{eq:Wetterich}
\partial_k \Gamma_k[\phi]=
\:\,\frac{1}{2}\,\Tr \left\{\partial_kR_k \left(\Gamma_k^{(2)}[\phi]+R_k\right)^{-1}\right\},
\end{equation}
where the $2$-point function $\Gamma_k^{(2)}[\phi]$ represents the second functional derivative w.r.t.~the field $\phi$, and the trace includes integration over the loop momentum as well as summation over all internal indices. The Wetterich equation has the simple diagrammatic representation shown in Fig.~\ref{fig:flow_gamma_simple}. 

Despite its simple one-loop structure, due to the full field dependence in the 
2-point function this functional integro-differential equation can usually not be solved without truncations. In order to describe phase transitions and critical phenomena it has been proven to be successful to use the so-called `derivative expansion' as an Ansatz for $\Gamma_k$, i.e.
\begin{align}
\label{eq:derivative_expansion}
\Gamma_k[\phi]=\int d^4x\left\{U_k(\phi)+\frac{1}{2}Z_k(\phi)(\partial_\mu \phi)^2+\frac{1}{8}Y_k(\phi) (\partial_\mu \phi^2)^2+\mathcal{O}(\partial^4)\right\},
\end{align}
were $U_k(\phi)$, $Z_k(\phi)$ and $Y_k(\phi)$ are the field and scale-dependent potential and derivative terms \cite{Tetradis:1993ts,Berges:2000ew}. The latter include wave-function renormalization factors. In many cases already the leading order of this derivative expansion, at which $Z_k(\phi)=1$ and $Y_k(\phi)=0$, can provide a reasonably good quantitative description of thermodynamics and critical phenomena, see, e.g., \cite{Litim2001,Braun:2009si}. This truncation is also known as the `local potential approximation' (LPA) and will be employed in the following.

\section{Flow Equations for the Quark-Meson Model}
\label{sec:QM}

As mentioned in the introduction, the aim of the present study is to elucidate the role of broken chiral symmetry and it restoration at high temperature and density in the spectral properties of light hadrons. As this remains very difficult in QCD proper, we resort to the QM model which represents a low-energy effective theory of QCD and is based on chiral symmetry \cite{Jungnickel:1995fp,Schaefer:2004en}. The elementary degrees of freedom are given by the two lightest quarks, i.e.~the up and down quark, as well as the three pions and the sigma meson. The QM model is renormalizable and falls into the same $O(4)$ universality class as expected for QCD with two flavors. 

\subsection{Flow of the effective action}

\begin{figure}
	\centerline{{\includegraphics[width=0.5\textwidth]{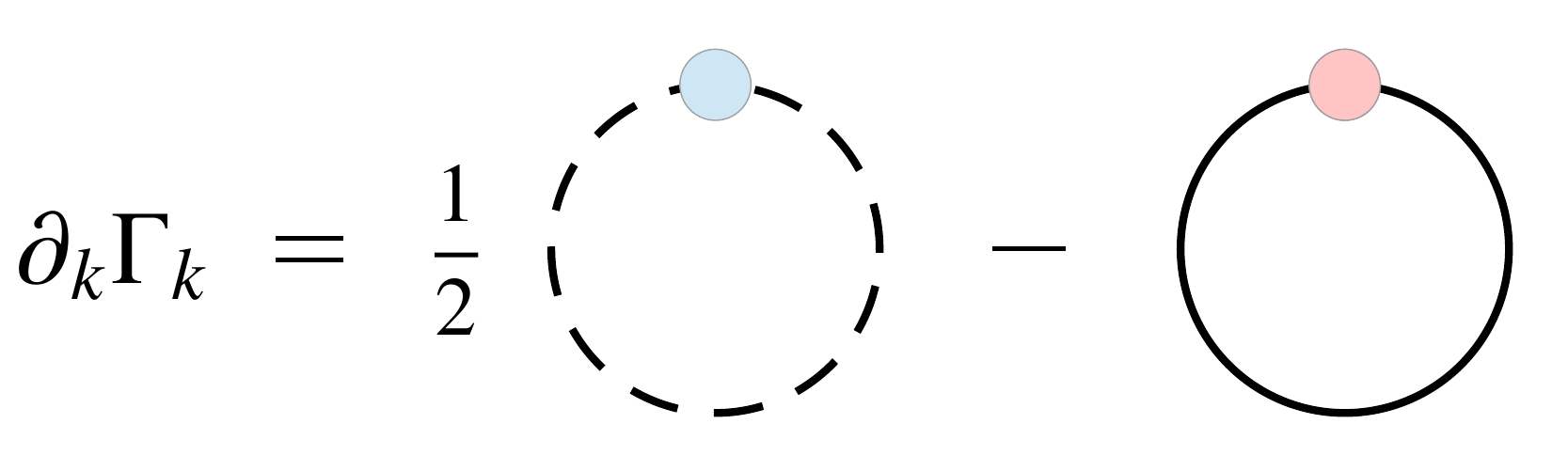}}}
	\caption{Diagrammatic representation of the flow equation for the effective average action of the quark-meson model. The dashed (solid) line represents the full scale-dependent mesonic (quark) propagator and the circles the regulator insertions $\partial_kR_k$.}
	\label{fig:flow_gamma}
\end{figure}

Applying the LPA yields the following Ansatz for the scale-dependent effective action of the quark-meson model,
\begin{equation}\label{eq:gamma}
\Gamma_{k}=
\int d^{4}x \:\left\{
\bar{\psi}\left(\gamma_\mu\partial^\mu+
h(\sigma+i\vec{\tau}\vec{\pi}\gamma^{5}) -\mu \gamma_0 \right)\psi
+\frac{1}{2} (\partial_{\mu}\phi)^{2}+U_{k}(\phi^2)-c\sigma
\right\}.
\end{equation}
The field $\phi=(\sigma,\vec{\pi})$ combines the mesonic fields, $h$ is the Yukawa coupling, $\mu$ the quark chemical potential and $-c\sigma$ an explicit chiral symmetry breaking term. The mesonic potential $U_{k}(\phi^2)$ as a function of the $O(4)$ invariant $\phi^2 = \sigma^2 + \vec\pi^2$ can in principle assume an arbitrary form and encodes the mesonic interactions. Inserting this Ansatz into the Wetterich equation gives the flow equation for $\Gamma_k$, depicted diagrammatically in Fig.~\ref{fig:flow_gamma}. It immediately reduces to a flow equation for the effective potential, because $U_{k}$ is the only scale-dependent quantity within the LPA.

\noindent
In order to solve the flow equation for $\Gamma_k$, we need to specify the shape of $U_{k}$ at the UV scale $k=\Lambda$. We choose a UV potential with a chirally symmetric minimum at $ \phi_0^2=0 $ of the form
\begin{equation}
\label{eq:pot_UV} 
U_\Lambda(\phi^{2}) =
\frac{1}{2}m_\Lambda^{2}\phi^{2} +
\frac{1}{4}\lambda_\Lambda(\phi^{2})^{2}.
\end{equation}
The explicit values for the parameters are listed in Tab.~\ref{tab:parameters}. The fermionic minus sign in the flow equation for the effective potential drives the minimum away from zero analogous to a negative index of refraction in a self-focusing non-linear medium and leads to the physical parameters listed in the table in the IR.

Moreover, we have to choose suitable bosonic and fermionic regulator functions $R_k$ \cite{Pawlowski:2015mlf}. In the following we adopt three-dimensional regulator functions that only regulate the spatial momenta but leave the energy component unaffected, cf. \cite{Pawlowski:2015mia} for an alternative approach involving 4D regulators. The choice of 3D regulators will allow us to perform the analytic continuation from imaginary to real energies on the level of the flow equations. For explicit expression of the regulator functions and the flow equations we refer to \cite{Tripolt2014, Tripolt2014a}.

\begin{table}[ht]
\tbl{Parameter set for the quark-meson model and obtained vacuum values for the particle masses and the pion decay constant in the IR. The UV scale is chosen to be $\Lambda=1$~GeV and the IR scale is $k_{\text{IR}}=40$~MeV.}
{\begin{tabular}{@{}cccccccc@{}} \toprule
$m_\Lambda/\Lambda$ & $\lambda_\Lambda$ & $c/\Lambda^3$ &  $h$ & $\sigma_0\equiv f_\pi$ & $m_\pi$ & $m_\sigma$ &  $m_\psi$   \\ \colrule
0.794  & 2.00 & 0.00175 & 3.2 &93.5 MeV & 138 MeV & 509 MeV & 299 MeV \\ \botrule
\end{tabular}
}
\label{tab:parameters}
\end{table}

The flow equation for $U_k(\phi^2)$ is solved numerically on a one-dimensional grid in field space, i.e., the field $\phi^2$ is discretized in the direction of the $\sigma$-field and the flow equation, which turns into a set of coupled ordinary differential equations, is solved on each grid point $\phi^2_i$. For numerical reasons, these flow equations are not integrated down to $k=0$ but only to $k_{\text{IR}}=40$~MeV which does, however, not affect our results since the minimum of the potential as well as the values of the particle masses are basically fixed as soon as the lightest degree of freedom decouples from the flow. 

In the IR, the explicit symmetry breaking term $c\sigma$ is subtracted from the effective potential and the resulting global minimum defines the pion decay constant as well as the curvature masses of the particles, cf.~Tab.~\ref{tab:parameters}. Moreover, the IR potential can be identified with the  Grand Canonical potential $\Omega (T,\mu)$ (Eq.~\ref{EffTP}) which allows access the various thermodynamic quantities such as pressure, entropy density and quark number density, 
\begin{align}
p(T,\mu)&=-\Omega (T,\mu)+\Omega (0,0),\\
s(T,\mu)&=\frac{\partial p(T,\mu)}{\partial T},\\
n_\psi(T,\mu)&=\frac{\partial p(T,\mu)}{\partial \mu}.
\end{align}

\subsection{Flow equations for spectral functions}

\begin{figure}
\centerline{\includegraphics[width=1.1\textwidth]{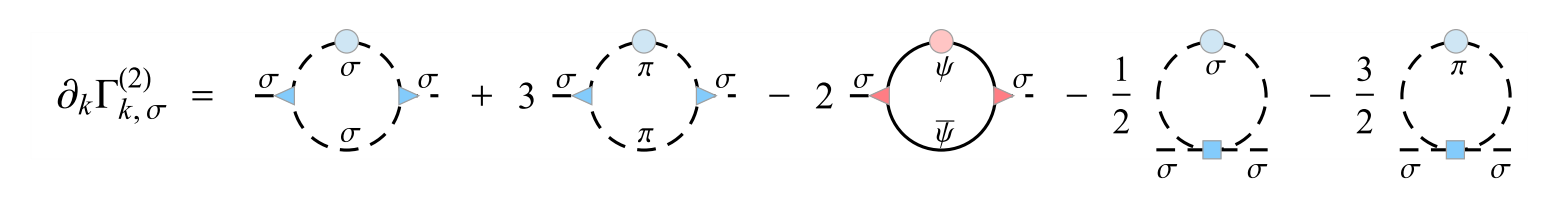}}	\centerline{\includegraphics[width=1.1\textwidth]{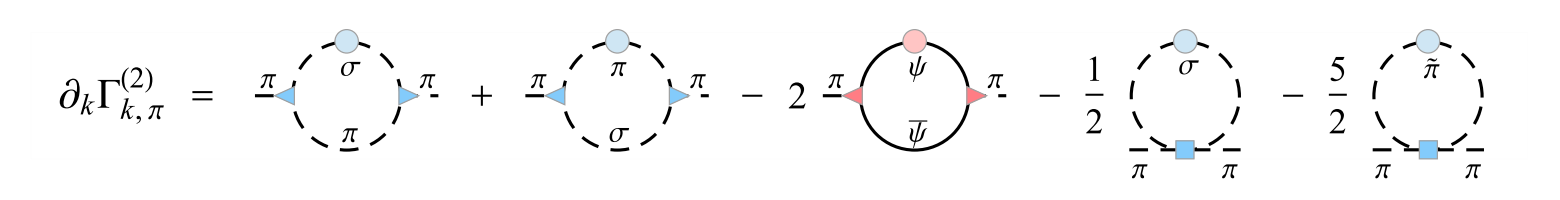}}
\centerline{\includegraphics[width=1.1\textwidth]{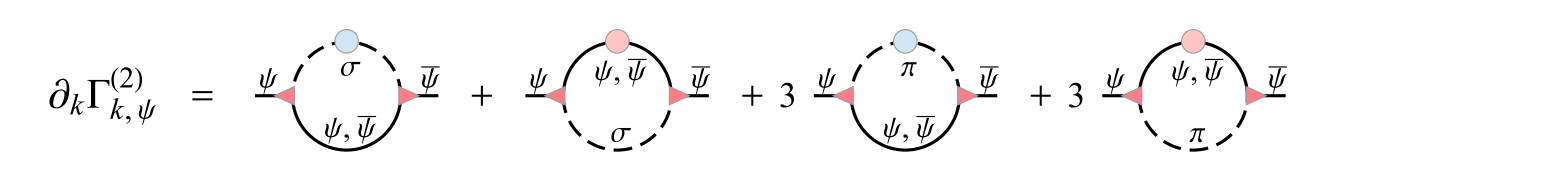}}
\caption{Diagrammatic representation of the flow equations for the sigma, pion and quark 2-point functions. Blue (red) triangles indicate the mesonic (quark-meson) three-point vertices, blue squares denote mesonic four-point vertices.}
\label{fig:flow_gamma2} 
\end{figure}

We now turn to the calculation of spectral functions which are derived from the retarded 2-point functions $\Gamma^{(2),R}$. The corresponding flow equations are obtained by taking two functional derivatives of the flow equation for the effective action, cf.~Fig.~\ref{fig:flow_gamma}, with respect to the appropriate fields. The structure of the resulting equations for the sigma, pion and quark 2-point functions is displayed in Fig.~\ref{fig:flow_gamma2}. 
The mesonic three- and four-point vertices are generated from derivatives of the scale-dependent effective potential $U_k$ and the quark-meson vertices are basically given by the Yukawa coupling. We note that this truncation is thermodynamically consistent in the sense that the (Euclidean) 2-point functions, evaluated at vanishing external momenta, agree with the curvature masses as extracted from the scale-dependent effective potential~\cite{Kamikado2014,Tripolt2014}. The truncation is also symmetry conserving in that the pion is massless in the chiral limit in the spontaneously broken phase, as required by the Goldstone theorem. 

To begin with, we are working in Euclidean space-time involving imaginary Matsubara frequencies. In order to be able to calculate real-time 2-point functions and spectral functions, we have to perform an analytic continuation from imaginary to real energies. This analytic continuation is performed on the level of the flow equations and is achieved by the following two-step procedure. First, the periodicity of the bosonic and fermionic occupation numbers with respect to the Euclidean Matsubara frequencies $p_0=i\,2\pi n T$ is exploited,
\begin{equation}
n_{B,F}(E+i p_0)\rightarrow n_{B,F}(E).
\end{equation}
In a second step, $p_0$ is replaced by a continuous real energy $\omega$,
\begin{equation}
\label{eq:continuation2}
\partial_k\Gamma^{(2),R}_k(\omega,\vec p)=-\lim_{\epsilon\to 0} \partial_k\Gamma^{(2),E}_k(p_0=-\I(\omega+\I\epsilon), \vec p),
\end{equation}
where the limit $\epsilon\to 0$ can be taken analytically for the imaginary part of the 2-point functions. This analytic continuation yields the physical Baym-Mermin boundary conditions \cite{Baym1961, Landsman1987}. The resulting flow equations for the real and the imaginary part of the retarded 2-point functions are then solved numerically and the spectral functions are obtained as
\begin{equation}
\rho(\omega,\vec p)=-\frac{1}{\pi}\text{Im}\frac{1}{\Gamma^{(2),R}(\omega,\vec p)}.
\end{equation}

\section{Vacuum Potential and Spectral Functions}
\label{sec:vacuum}

\begin{figure}
	\centerline{
		{\includegraphics[width=0.5\textwidth]{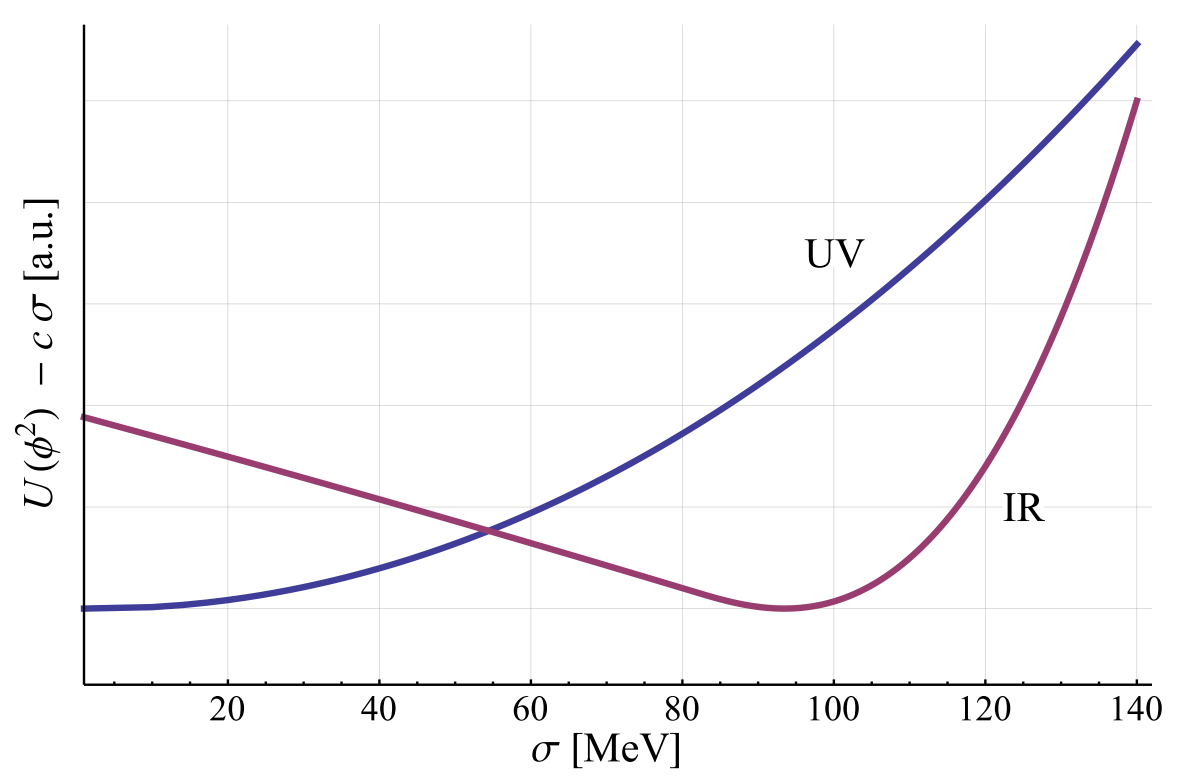}}	{\includegraphics[width=0.5\textwidth]{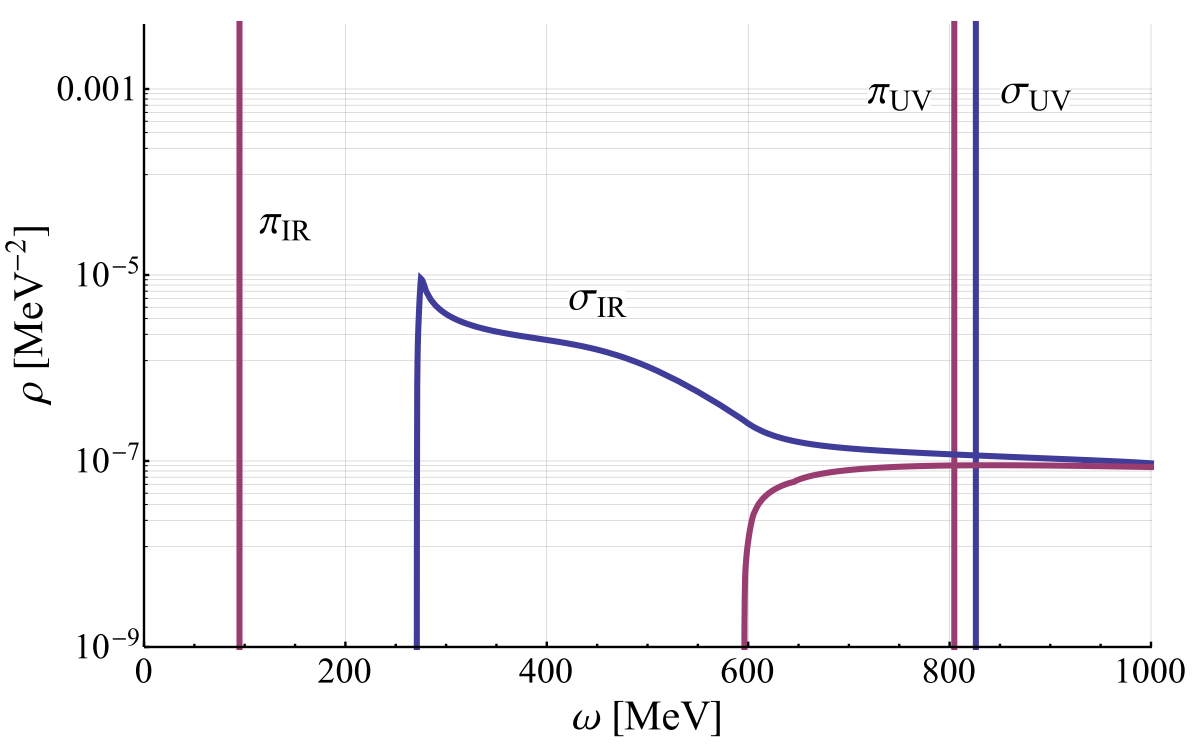}}
	}
	\caption{Left: The effective potential $U_k(\phi^2)-c\sigma_k$ is shown in the UV ($k_\Lambda)$ and the IR
	$(k=0)$. Right: The sigma- and pion spectral functions are shown in the UV and the IR.}
	\label{fig:IR_UV}
\end{figure}

To elucidate the relevant physical effects generated by the flow with resolution scale $k$, we first present results for the flow of the effective potential $U_k$ and the pion- and sigma spectral functions in the vacuum, see also \cite{Kamikado2014, Tripolt2014}. As shown in Fig.~\ref{fig:IR_UV}, the UV shape of the effective potential is essentially symmetric, up to a small effect due to the explicit symmetry breaking term $c\sigma$. When solving the corresponding flow equation, cf.~Fig.~\ref{fig:flow_gamma}, with decreasing $k$ the quark fluctuations push the minimum of the effective potential to larger values of the sigma field, thus establishing the well-known Mexican-hat shape of the chirally broken potential.

In Fig.~\ref{fig:IR_UV}, we also show the vacuum sigma and pion spectral functions at the UV and the IR scale. As a result of the quasi-symmetric UV potential, the UV spectral functions are almost degenerate with masses of the order of 800 MeV. Both the sigma meson and the pion are stable in the UV since all available decay channels are essentially boosted to very high energies due to the large scale $k$ which can be loosely interpreted as the three-momentum $|\vec{p}|$. In the IR, however, the sigma meson becomes unstable with a large width, since it can decay into two pions,. The pion as a Goldstone particle remains stable. These results nicely illustrate the transition from a linear realization of chiral symmetry to a non-linear one through the RG-flow.

\section{Phase Diagram and Thermodynamics}
\label{sec:thermo}

\begin{figure}
	\centerline{
		{\hspace{-40mm}\includegraphics[width=0.62\textwidth]{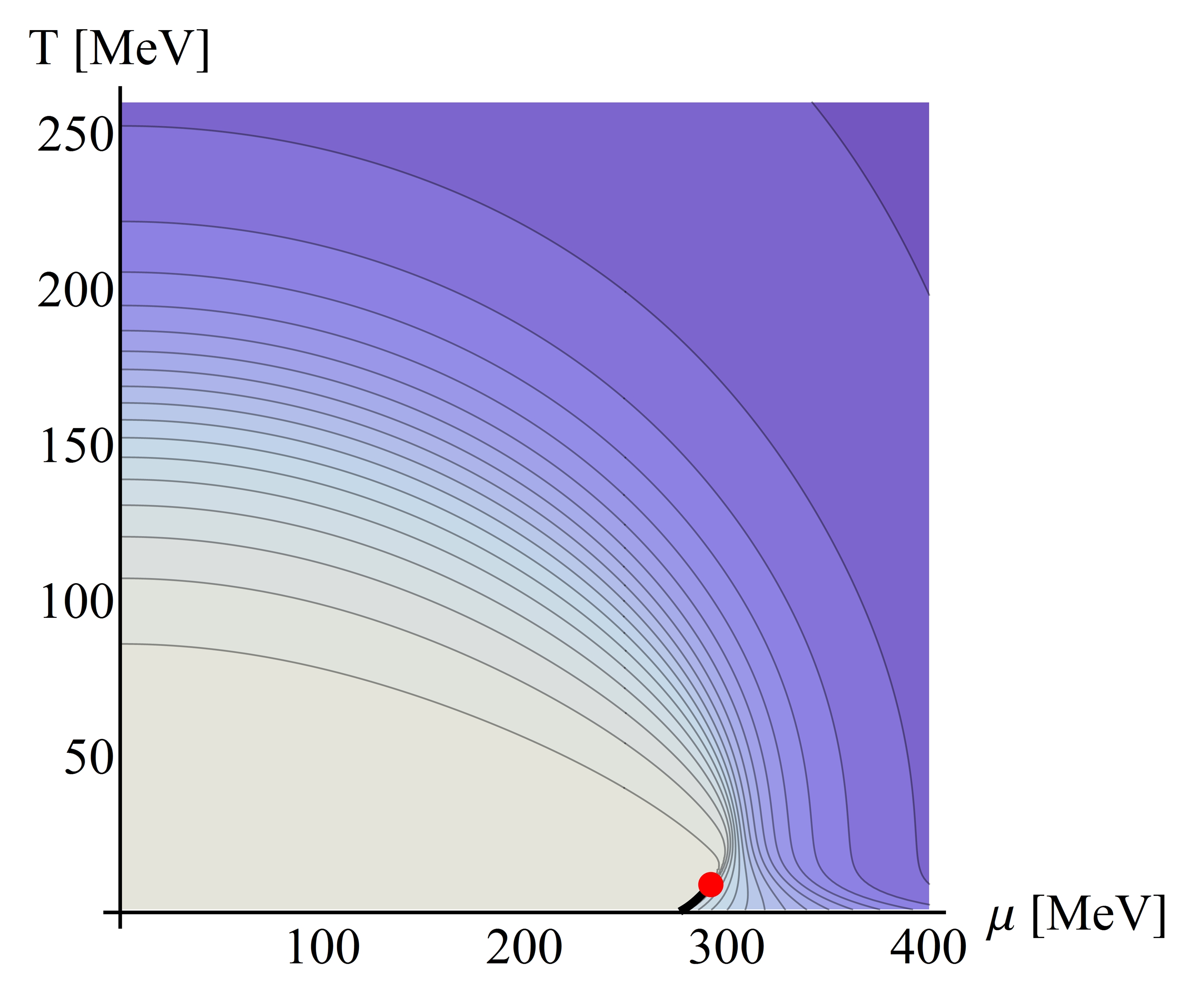}\llap{\makebox[1.0cm][l]{\raisebox{0.9cm}{\includegraphics[height=4.2cm]{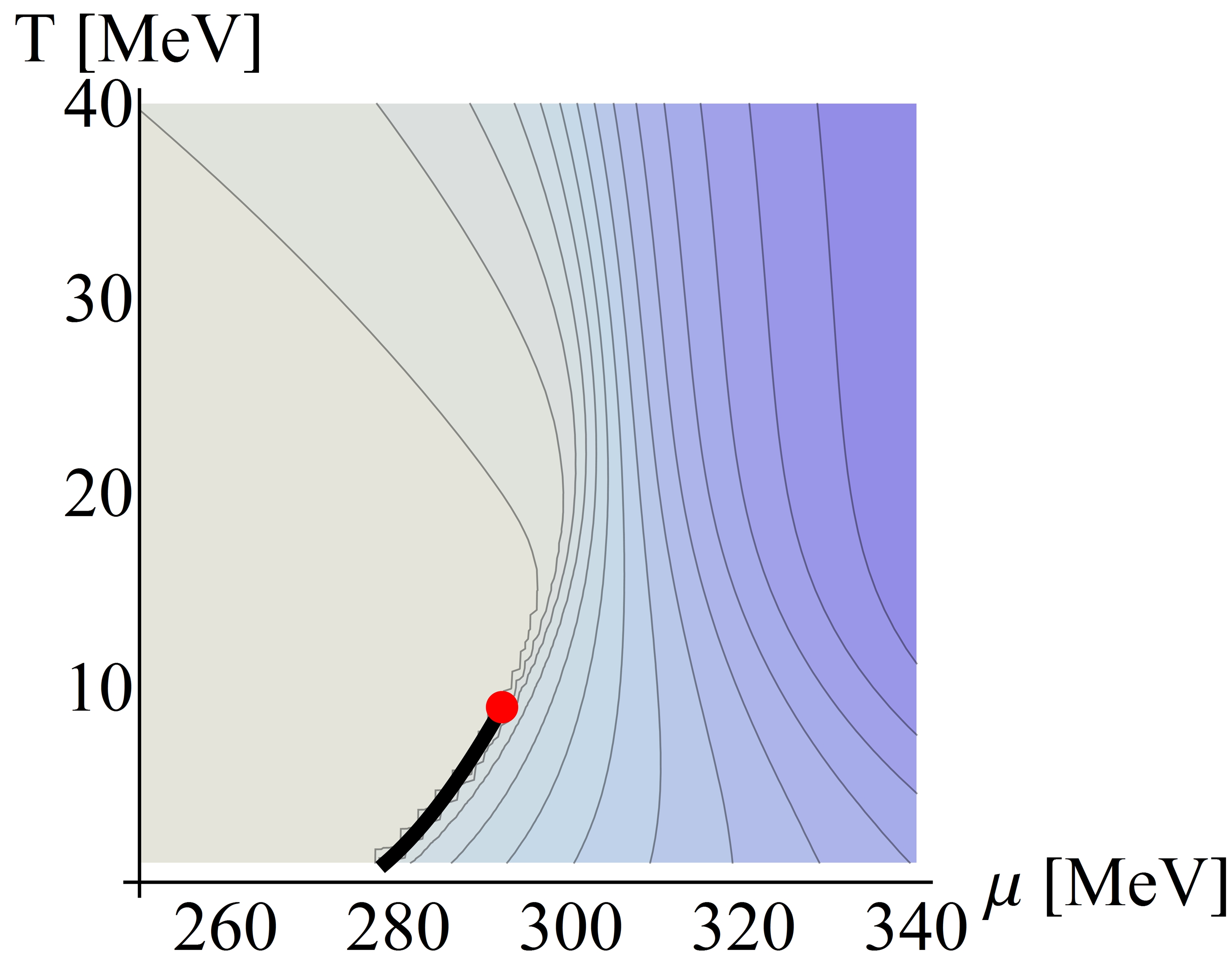}}}}}
		}
	\caption{The phase diagram of the quark-meson model is shown as a contour plot of the IR chiral order parameter $\sigma_{k=0}$ which decreases towards higher temperatures and larger quark chemical potentials, cf.~\cite{Tripolt2014}. The red dot represents the critical endpoint and the black line the first-order phase transition. Right: Enlargement of the regime around the CEP.}
	\label{fig:phase_diagram}
\end{figure}

In this section we present results for the phase diagram of the QM model and other thermodynamical quantities. The phase diagram, as obtained from the location of the global minimum of the effective potential, $\sigma_{k=0}(\mu,T)$, is shown in Fig.~\ref{fig:phase_diagram}. 

It exhibits a chiral crossover transition, which is located at $T\approx 170$~MeV for $\mu=0$, a chiral critical endpoint (CEP) at $T_c\approx 9$~MeV and $\mu\approx 292$~MeV and a first-order phase transition for $T< T_c$. 

\begin{figure}
	\centerline{
		{\includegraphics[width=0.5\textwidth]{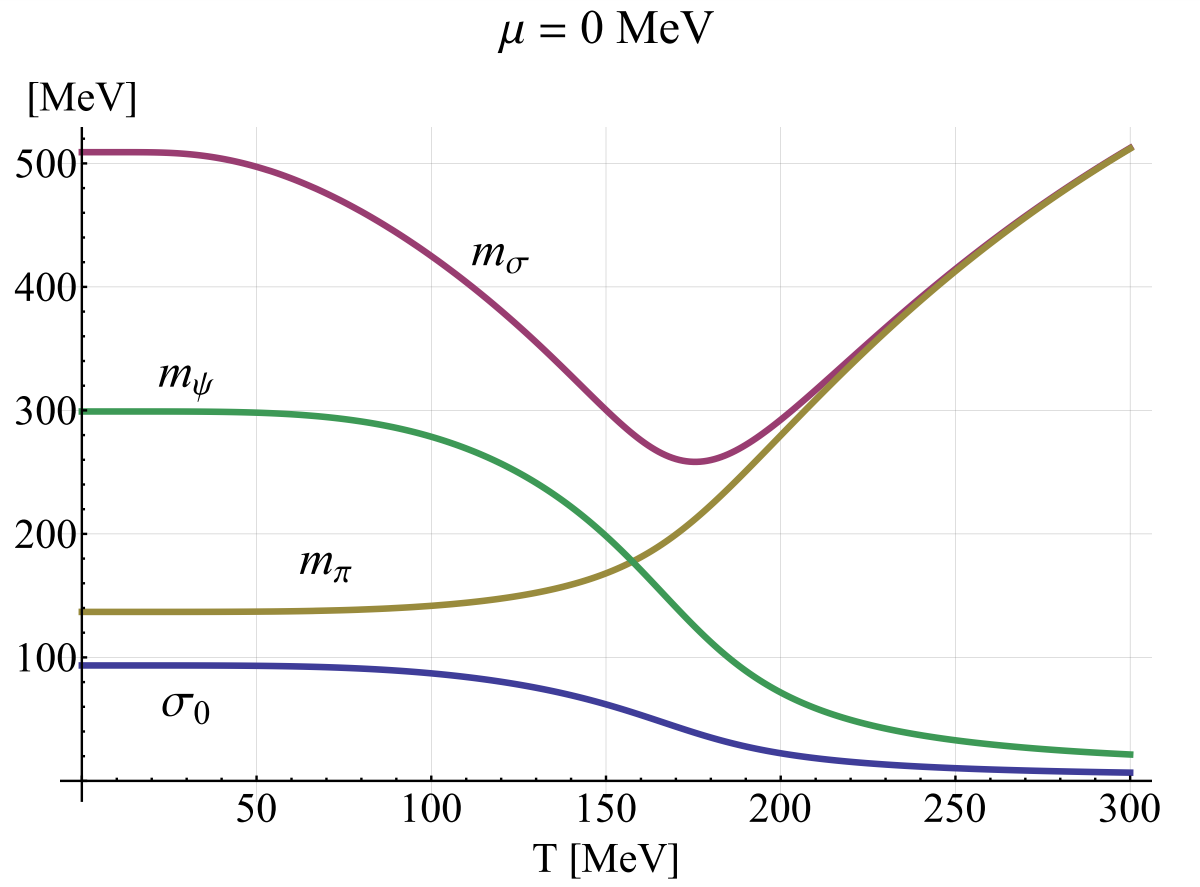}}	{\includegraphics[width=0.5\textwidth]{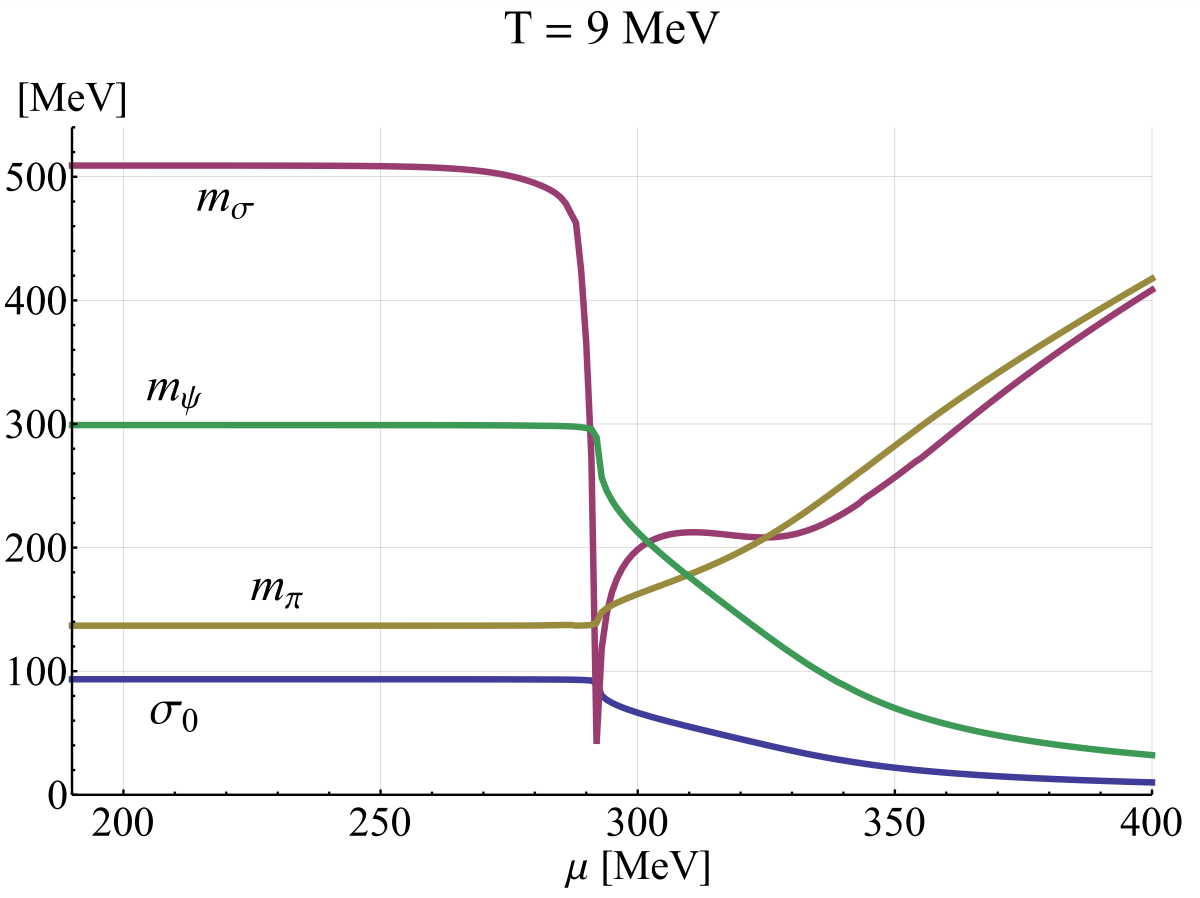}}
	}
	\caption{The IR curvature (screening) masses are shown together with the chiral order parameter $\sigma_0$ vs.~temperature at $\mu=0$ (left) and vs.~chemical potential at $T=9$~MeV (right), cf.~\cite{Tripolt2014}.}
	\label{fig:masses}
\end{figure}

To study the medium effects we will focus on two regions in the phase diagram - the temperature axis at $\mu=0$ and a horizontal line at $T\approx 9$~MeV near the CEP. In Fig.~\ref{fig:masses} we show the behavior of the chiral order parameter and the curvature (screening) masses within these two regions of the phase diagram. Near the crossover transition, the sigma mass decreases and then becomes degenerate with the pion mass as chiral symmetry gets restored at high temperatures. At the same time, the quark mass and the chiral order parameter (identified with the weak pion decay constant), decrease. Near the CEP, which represents a second-order phase transition, the sigma meson is expected to become massless, which is reproduced by our numerical calculations to a very good degree. At higher chemical potentials, we again observe a progressing degeneration of the sigma meson and the pion.


\begin{figure}
	\centerline{
		{\includegraphics[width=0.5\textwidth]{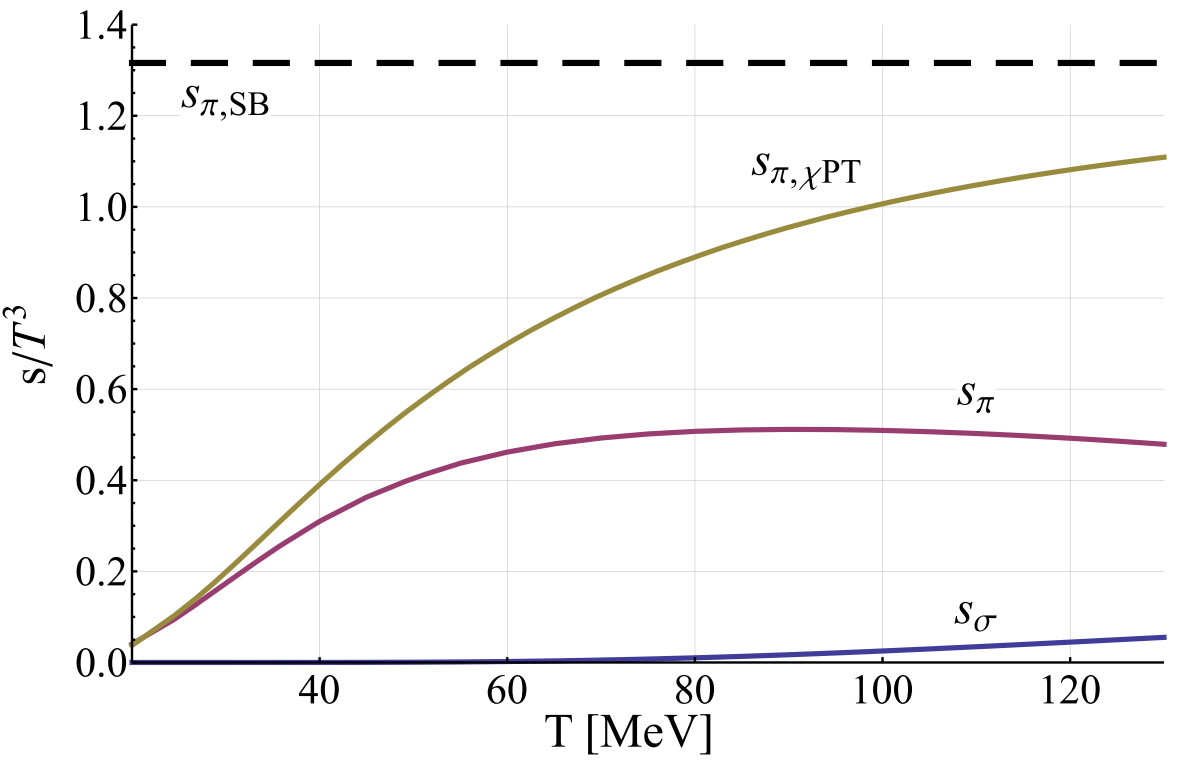}}
		{\includegraphics[width=0.5\textwidth]{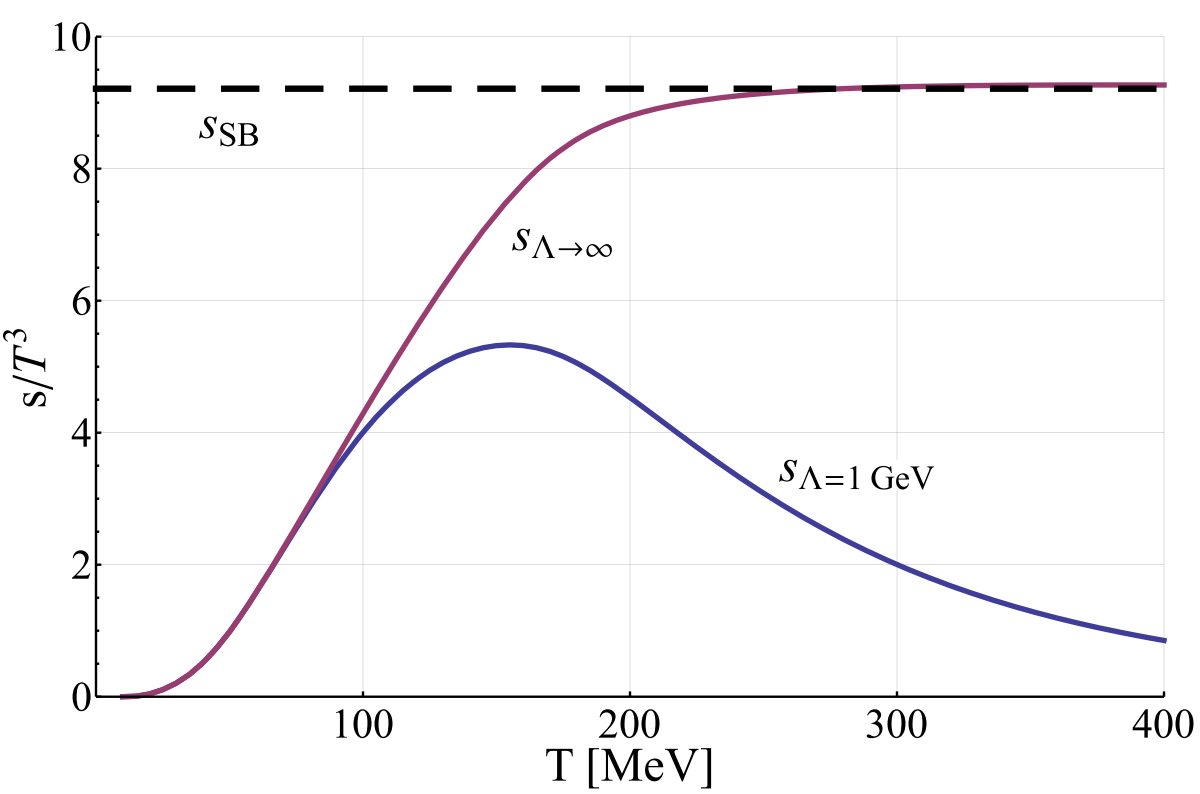}}
	}
	\caption{The entropy density is shown for the pions and the sigma mesons vs.~temperature at $\mu=0$ in comparison to the a result from $\chi$PT, \cite{Lang2012}, and the SB value for three massless non-interacting pions of $s_{\pi,\text{SB}}/T^3=2\pi^2/15$ (dashed line). Right: The total entropy density of the quark-meson model is shown vs.~temperature at $\mu=0$ without UV corrections, $s_{\Lambda=1\,\text{GeV}}$, and with UV corrections, $s_{\Lambda\rightarrow \infty}$, in comparison to the corresponding SB value of $s_{\text{SB}}/T^3=14\pi^2/15$ (dashed line).}
	\label{fig:entropy}
\end{figure}

For later discussions we show in Fig.~\ref{fig:entropy} the entropy density of the mesons and of the complete quark-meson system. The entropy density of the pions agrees well with the chiral perturbation theory ($\chi$PT) result obtained in \cite{Lang2012} at low temperatures. The total entropy density is dominated by the quarks and approaches the corresponding Stefan-Boltzmann (SB) value at high temperatures once effects from scales beyond the UV cutoff $\Lambda$ are taken into account, see \cite{Braun:2003ii,Herbst:2010rf,Skokov:2010wb,Tripolt2015} for details.

\section{In-Medium Spectral Functions}
\label{sec:spectral}

\begin{figure}[ht]
\centerline{
	\includegraphics[width=0.17\columnwidth]{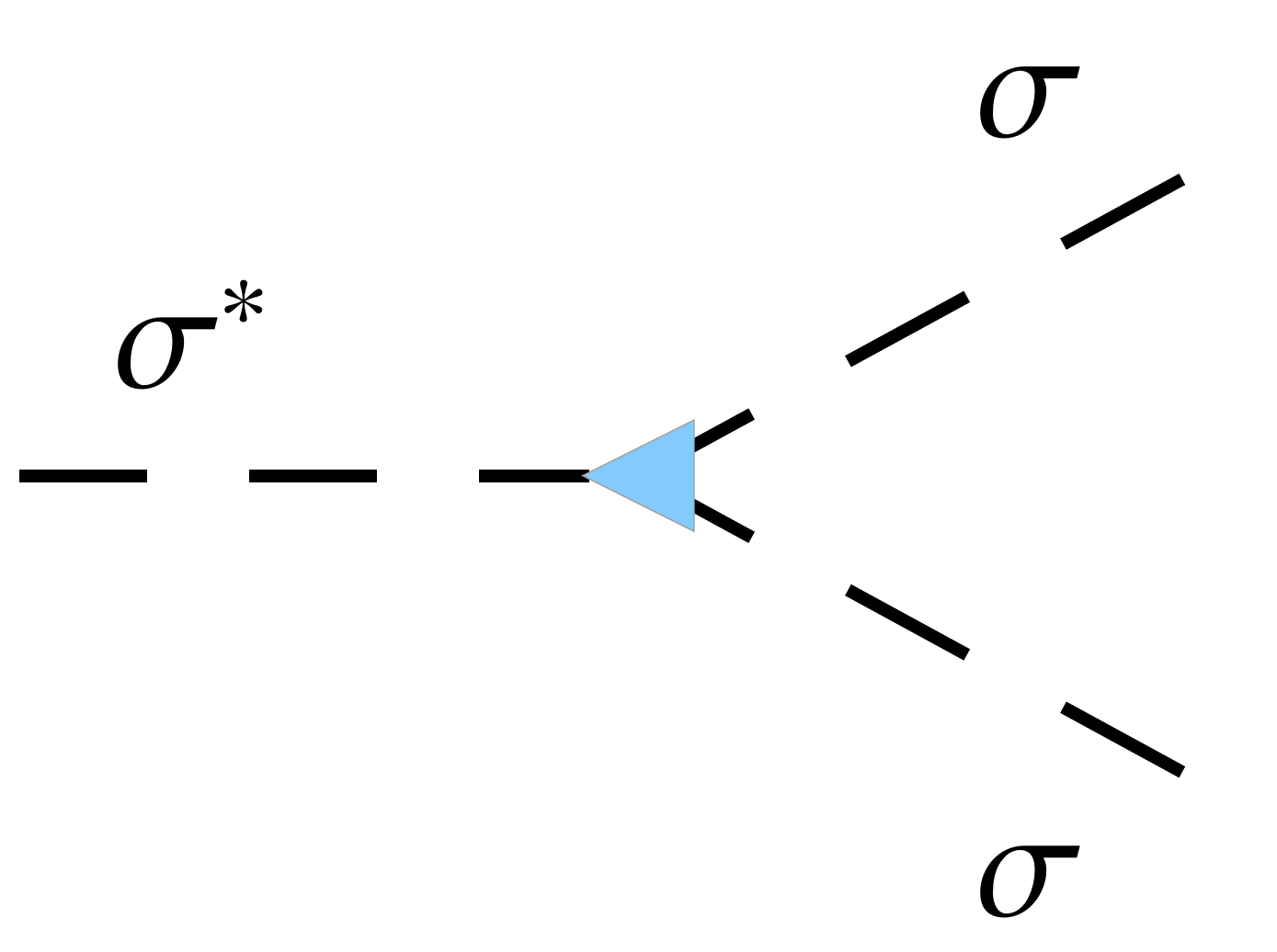}\hspace{8mm}
	\includegraphics[width=0.17\columnwidth]{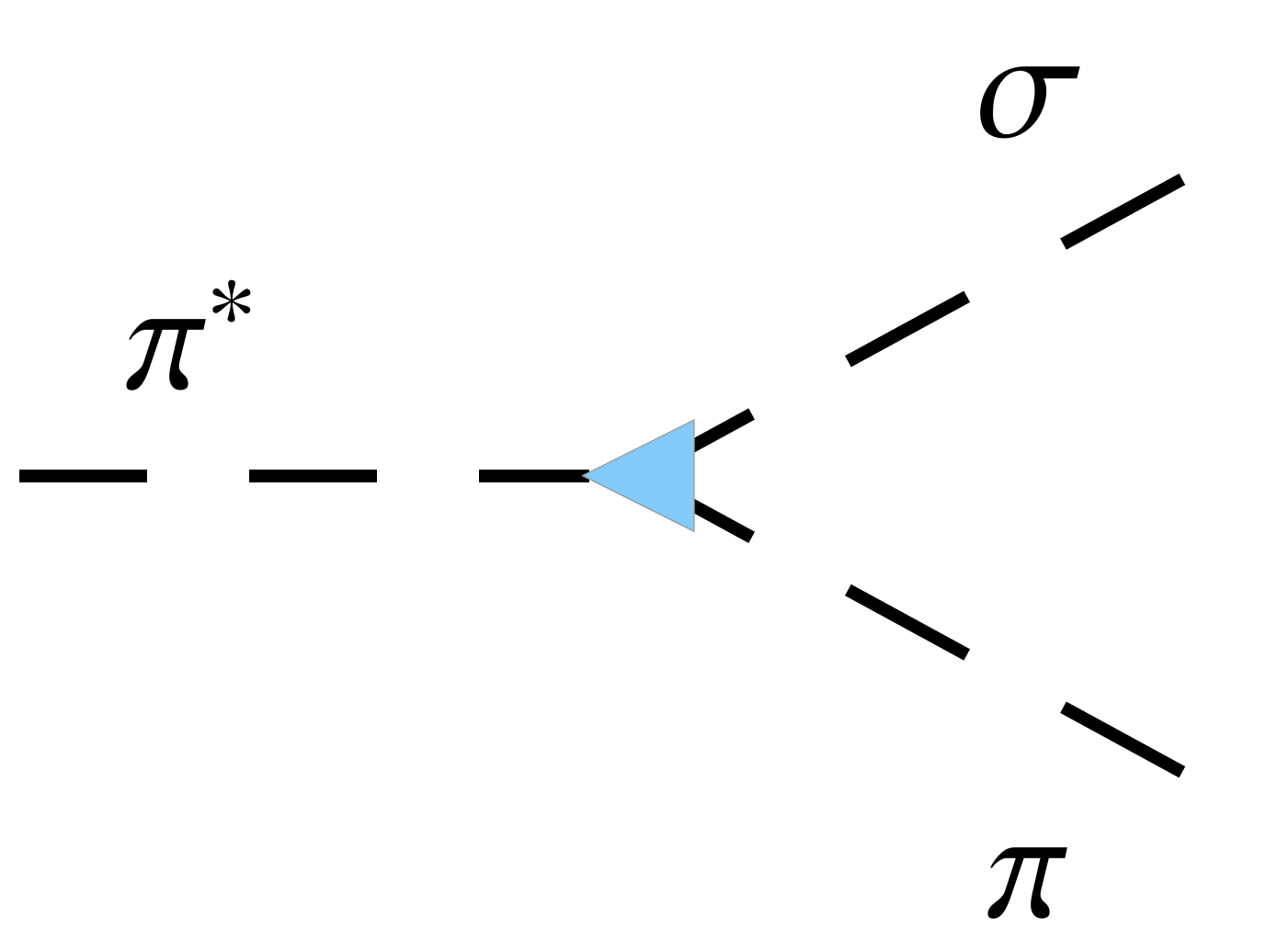}\hspace{8mm}
	\includegraphics[width=0.17\columnwidth]{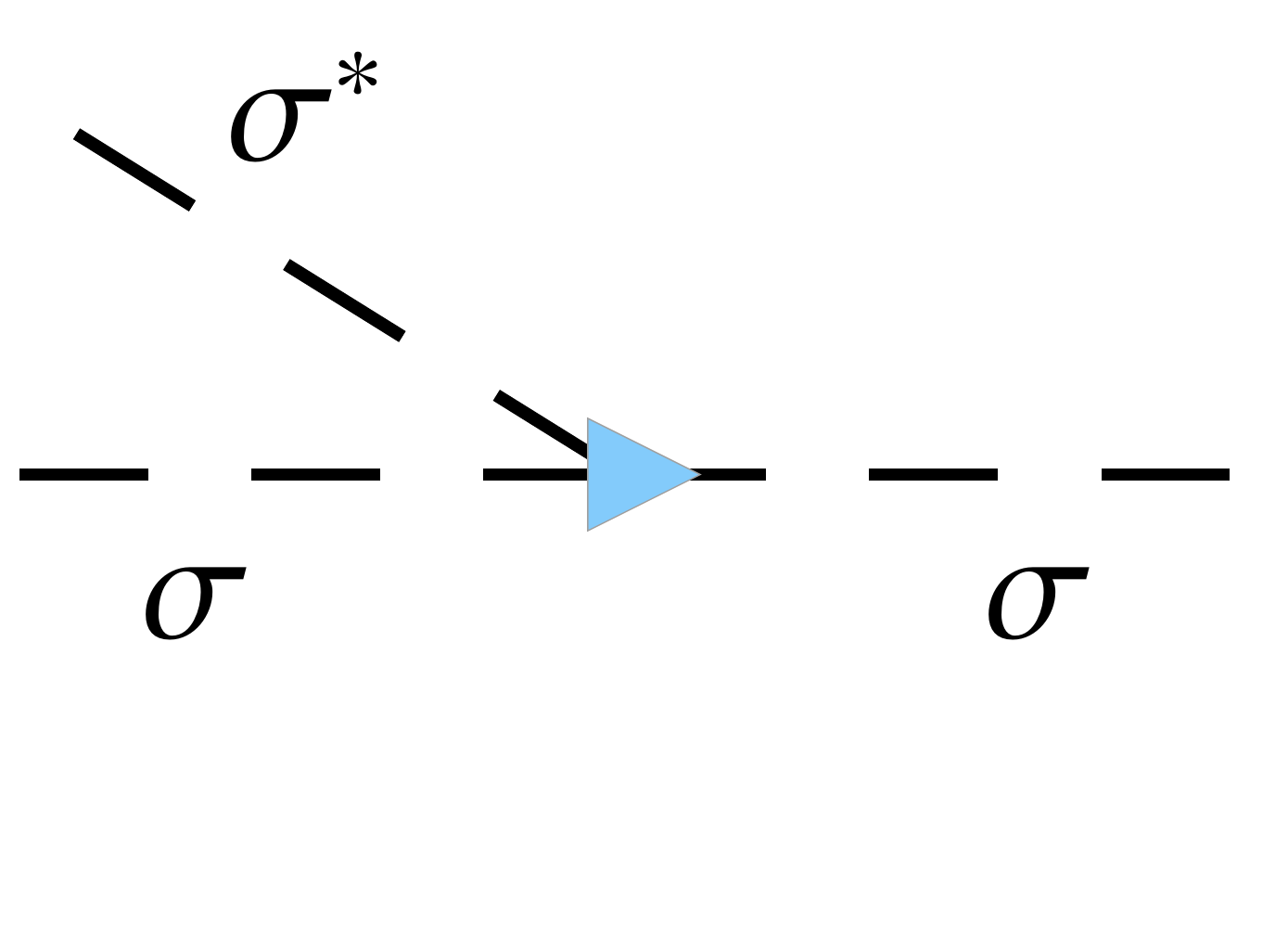}\hspace{8mm}
	\includegraphics[width=0.17\columnwidth]{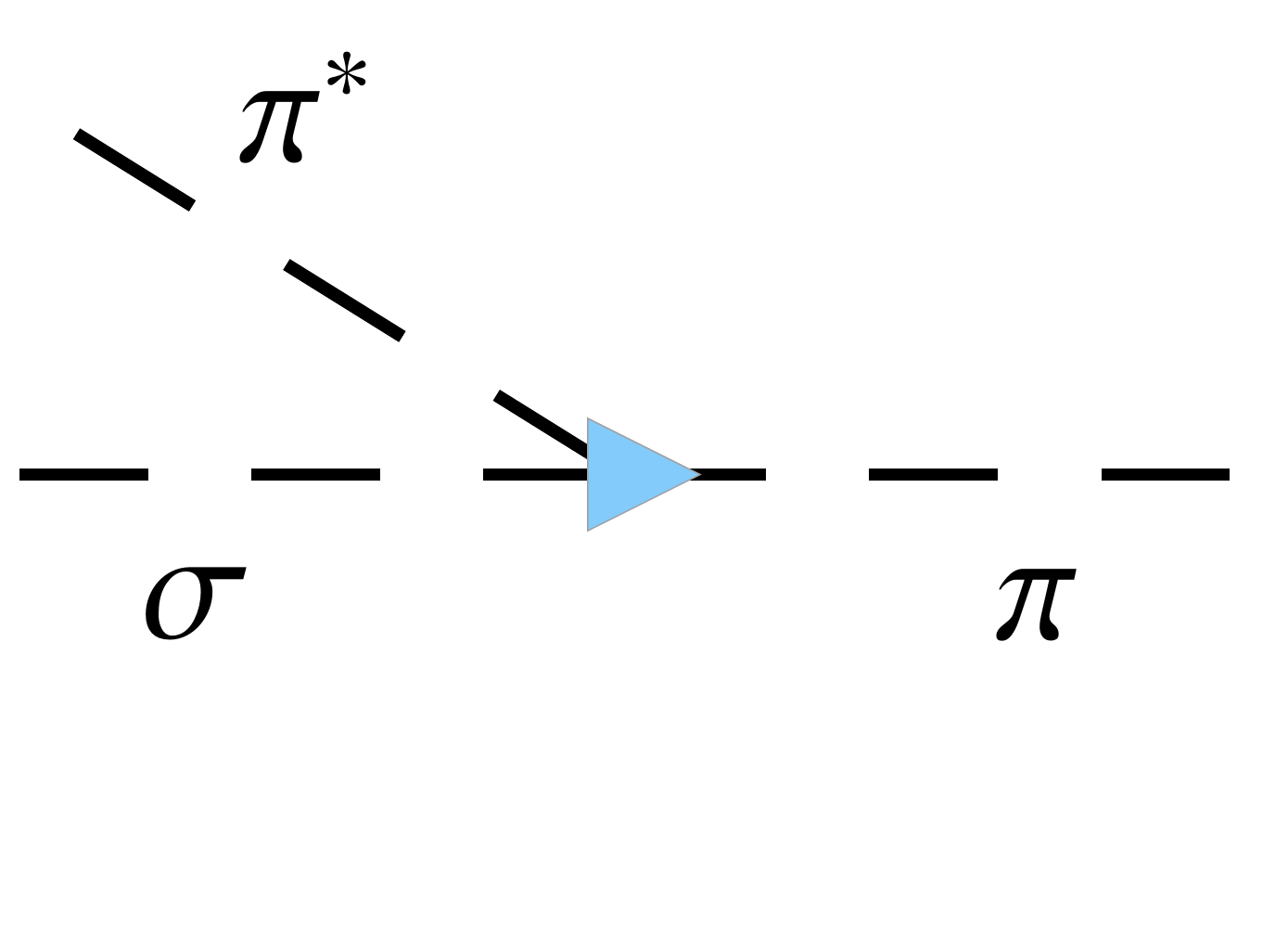}
}
\centerline{
	\includegraphics[width=0.17\columnwidth]{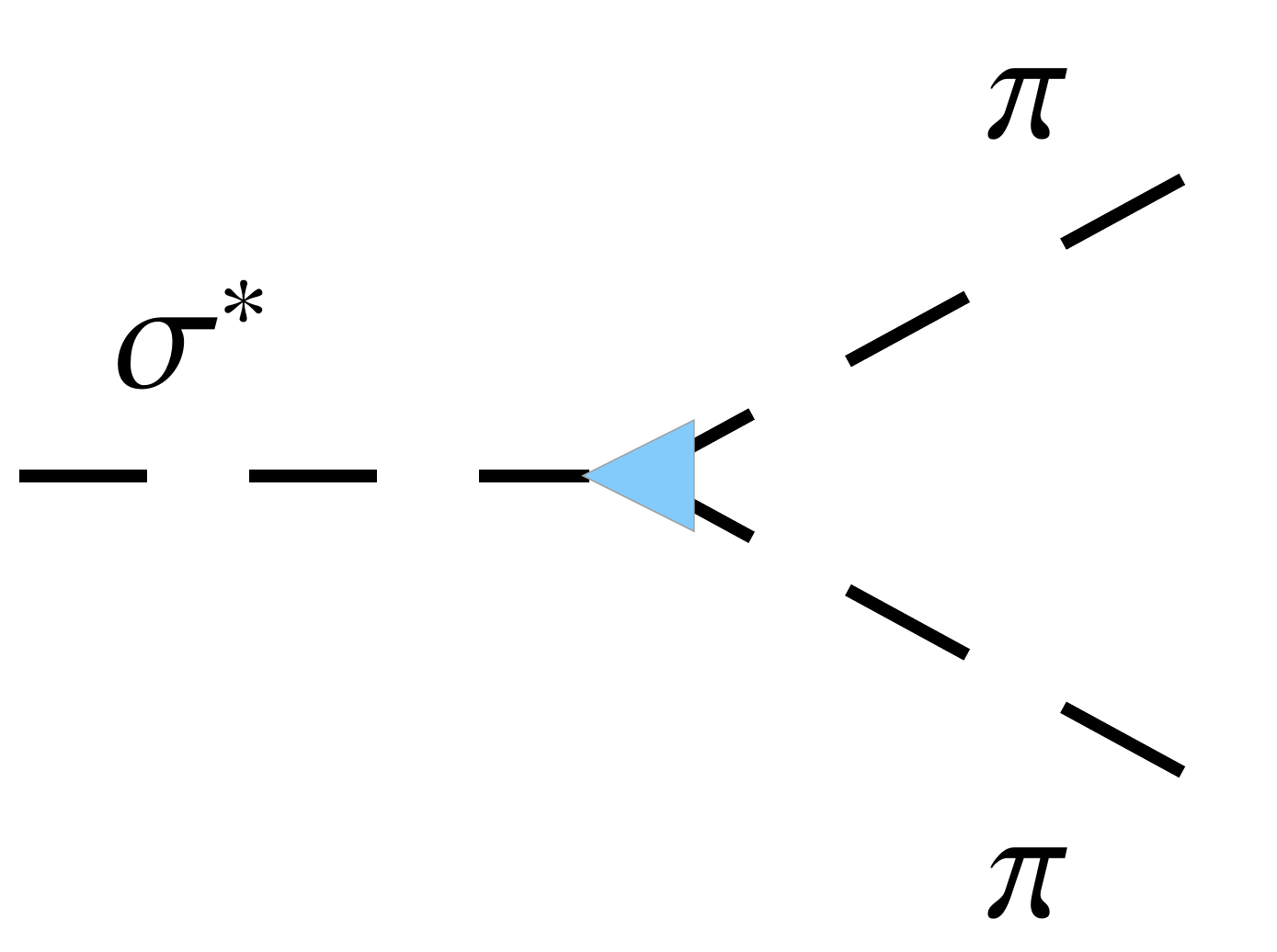}\hspace{8mm}
	\includegraphics[width=0.17\columnwidth]{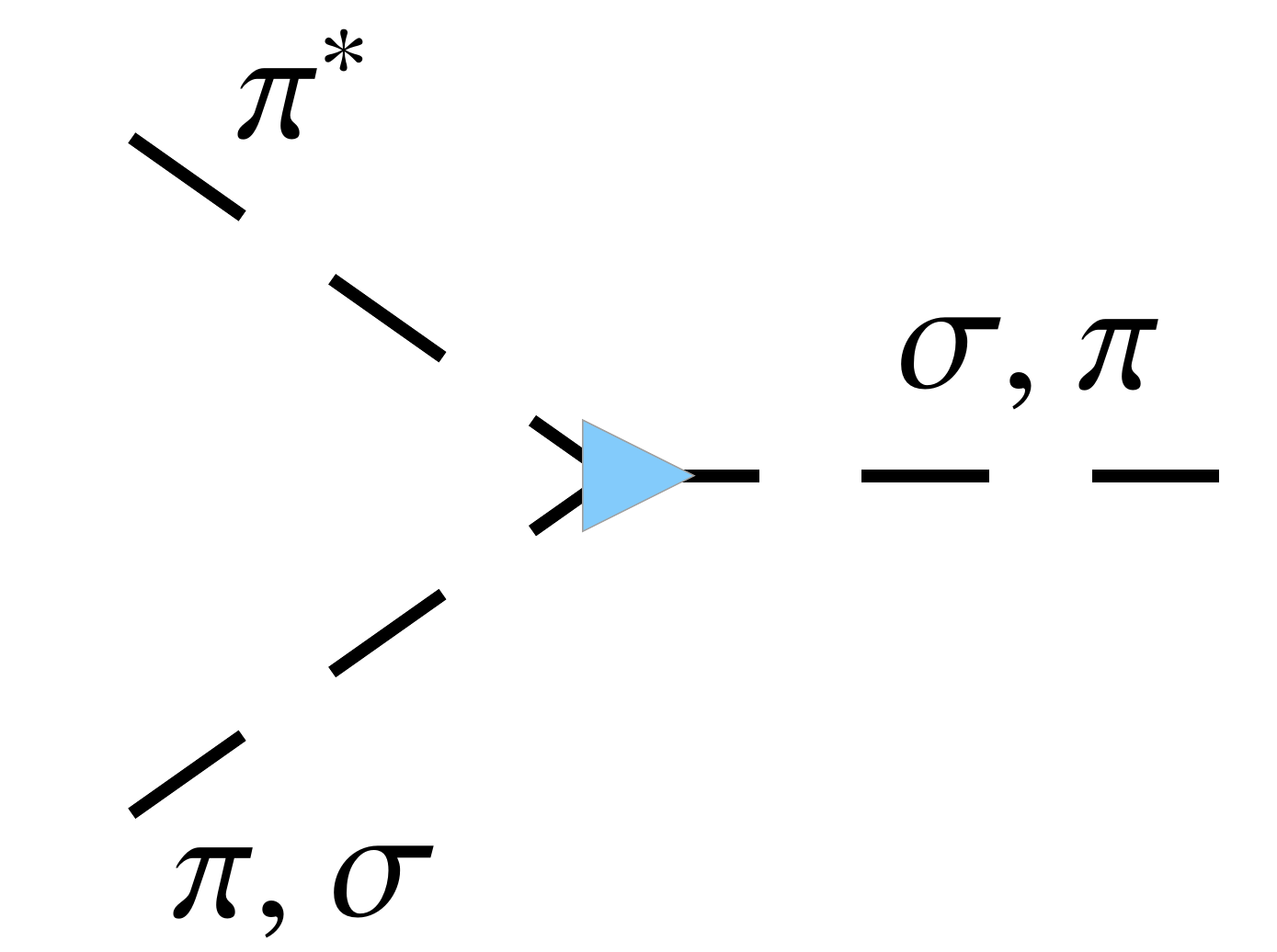}\hspace{8mm}
	\includegraphics[width=0.17\columnwidth]{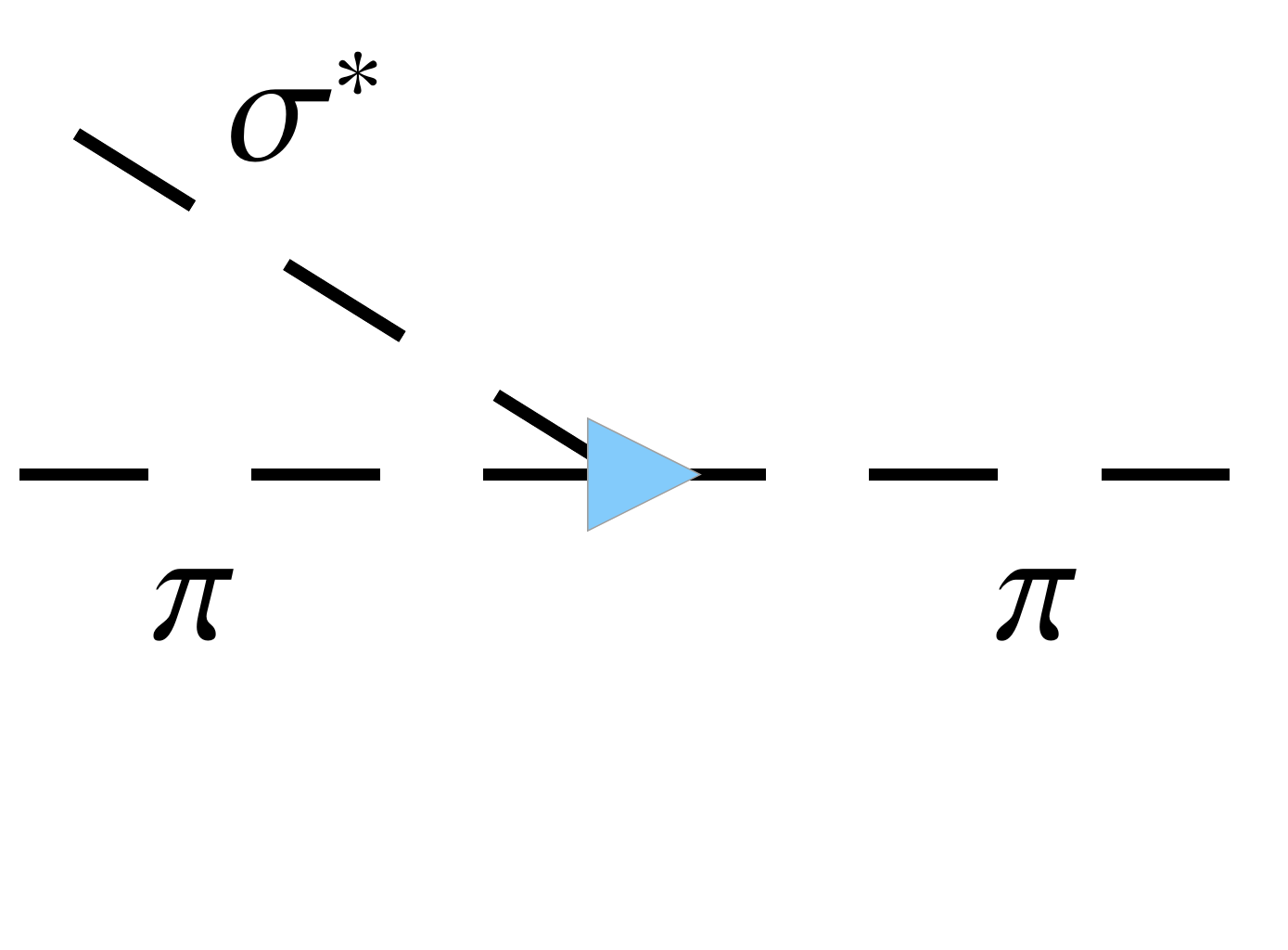}\hspace{8mm}
	\includegraphics[width=0.17\columnwidth]{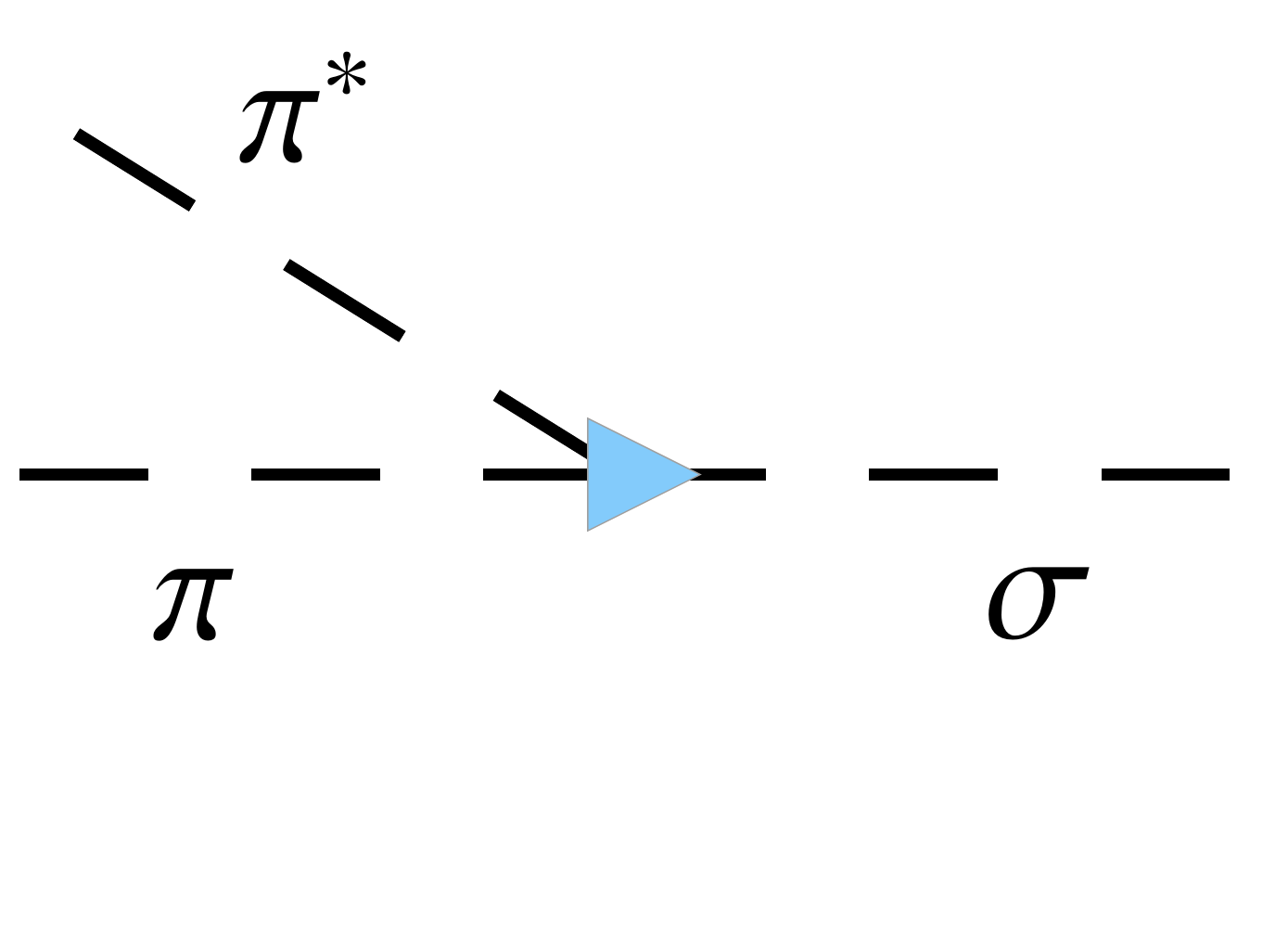}
}
\centerline{
	\includegraphics[width=0.17\columnwidth]{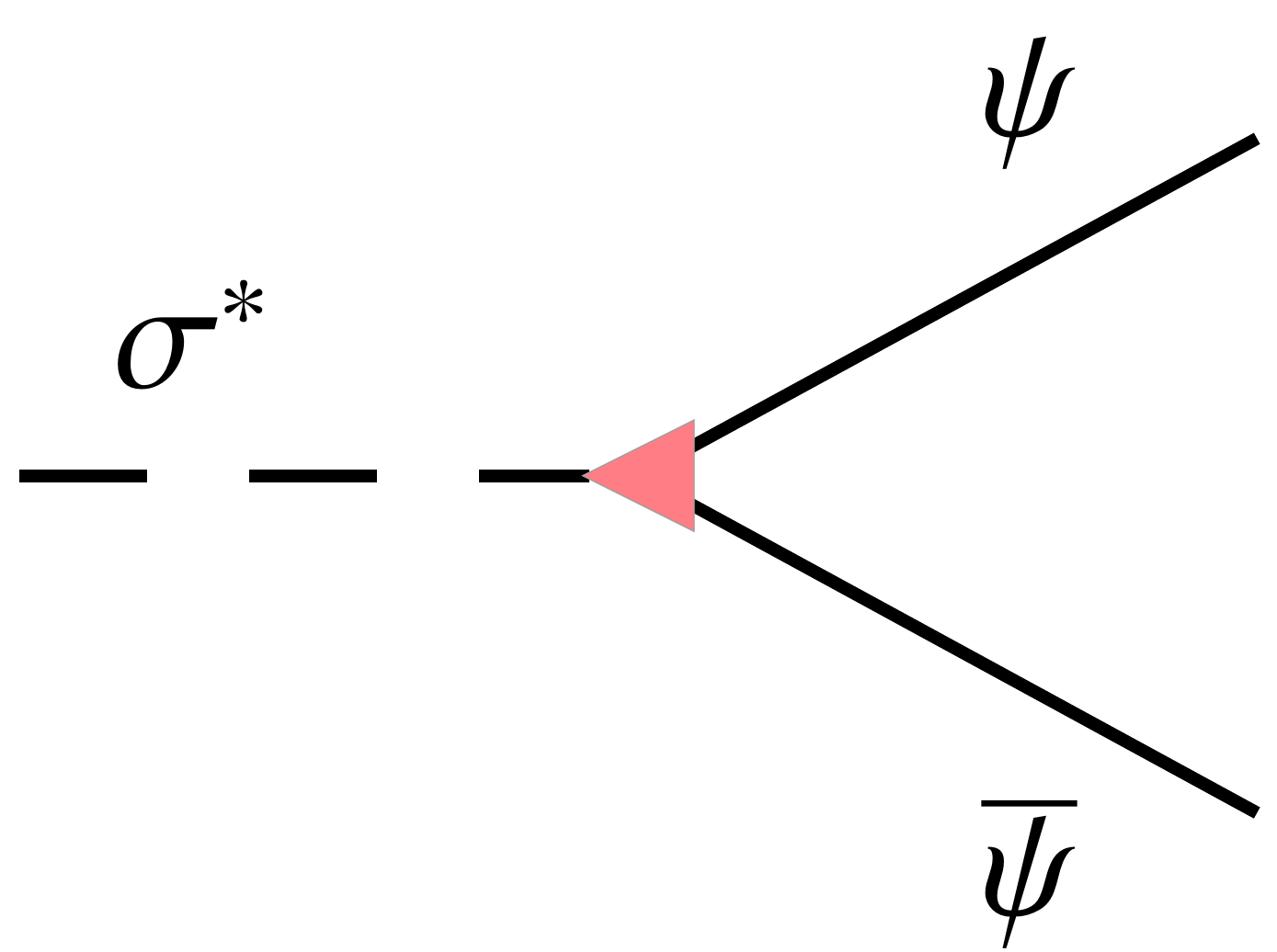}\hspace{8mm}
	\includegraphics[width=0.17\columnwidth]{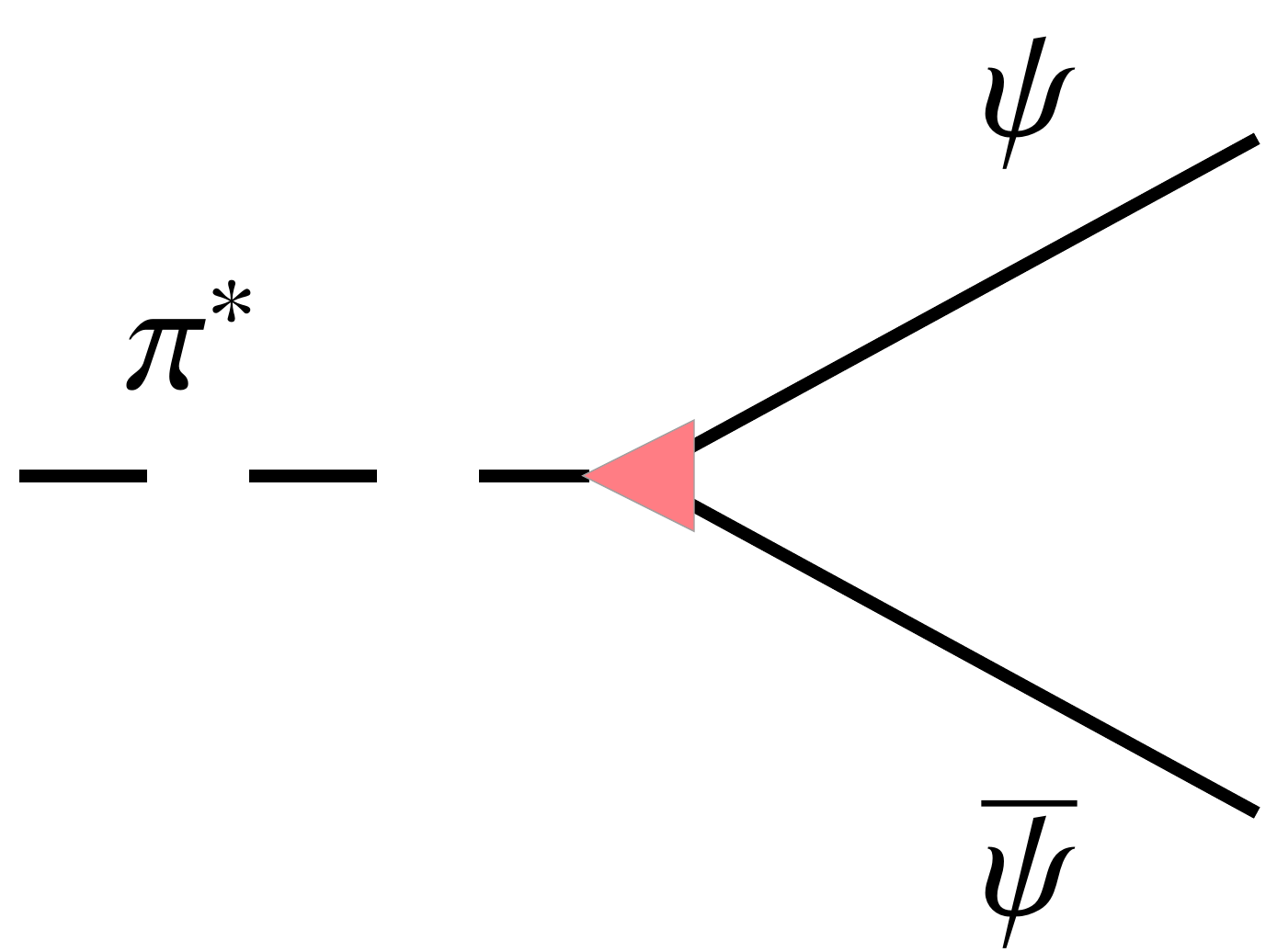}\hspace{8mm}
	\includegraphics[width=0.17\columnwidth]{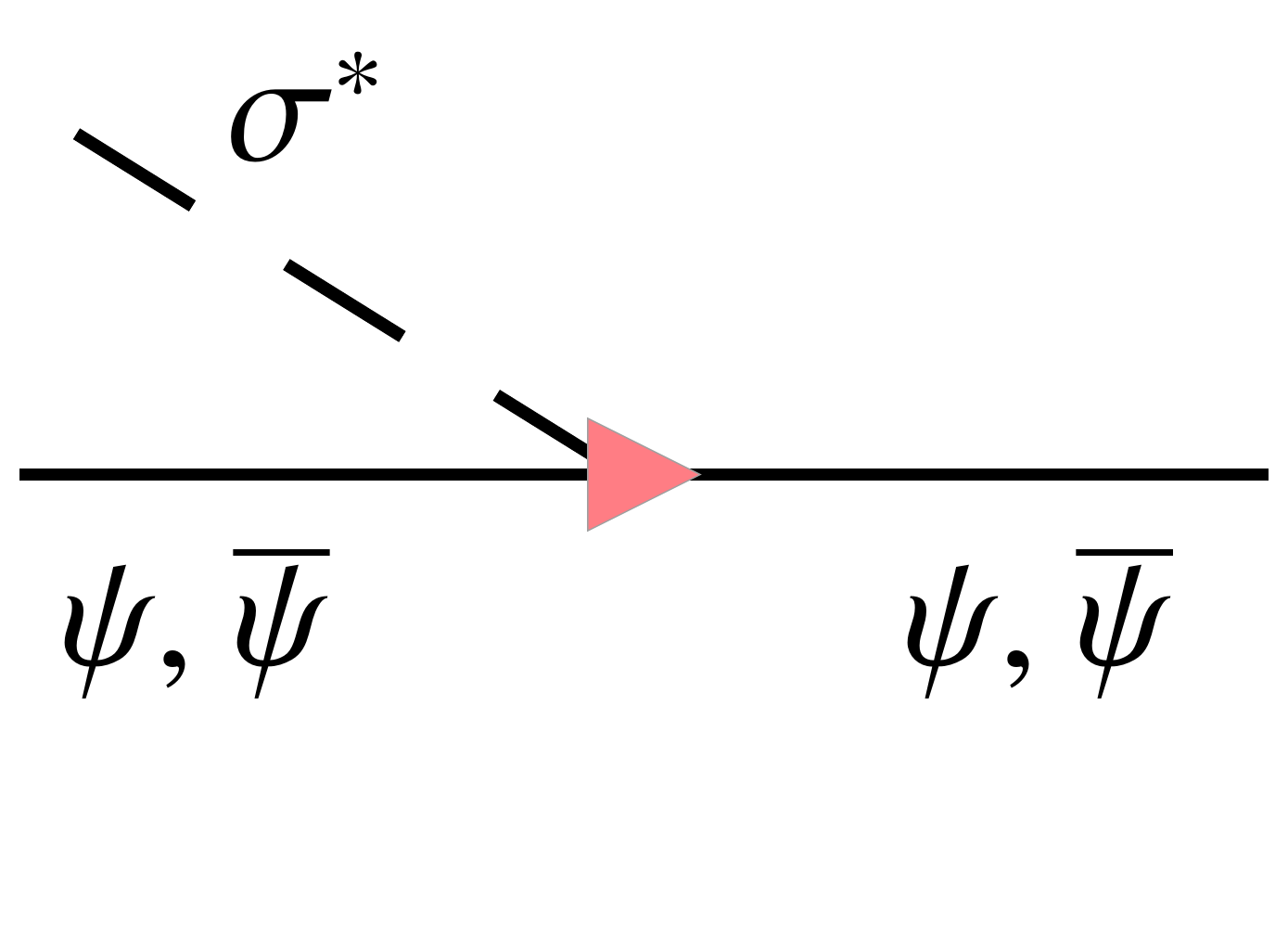}\hspace{8mm}
	\includegraphics[width=0.17\columnwidth]{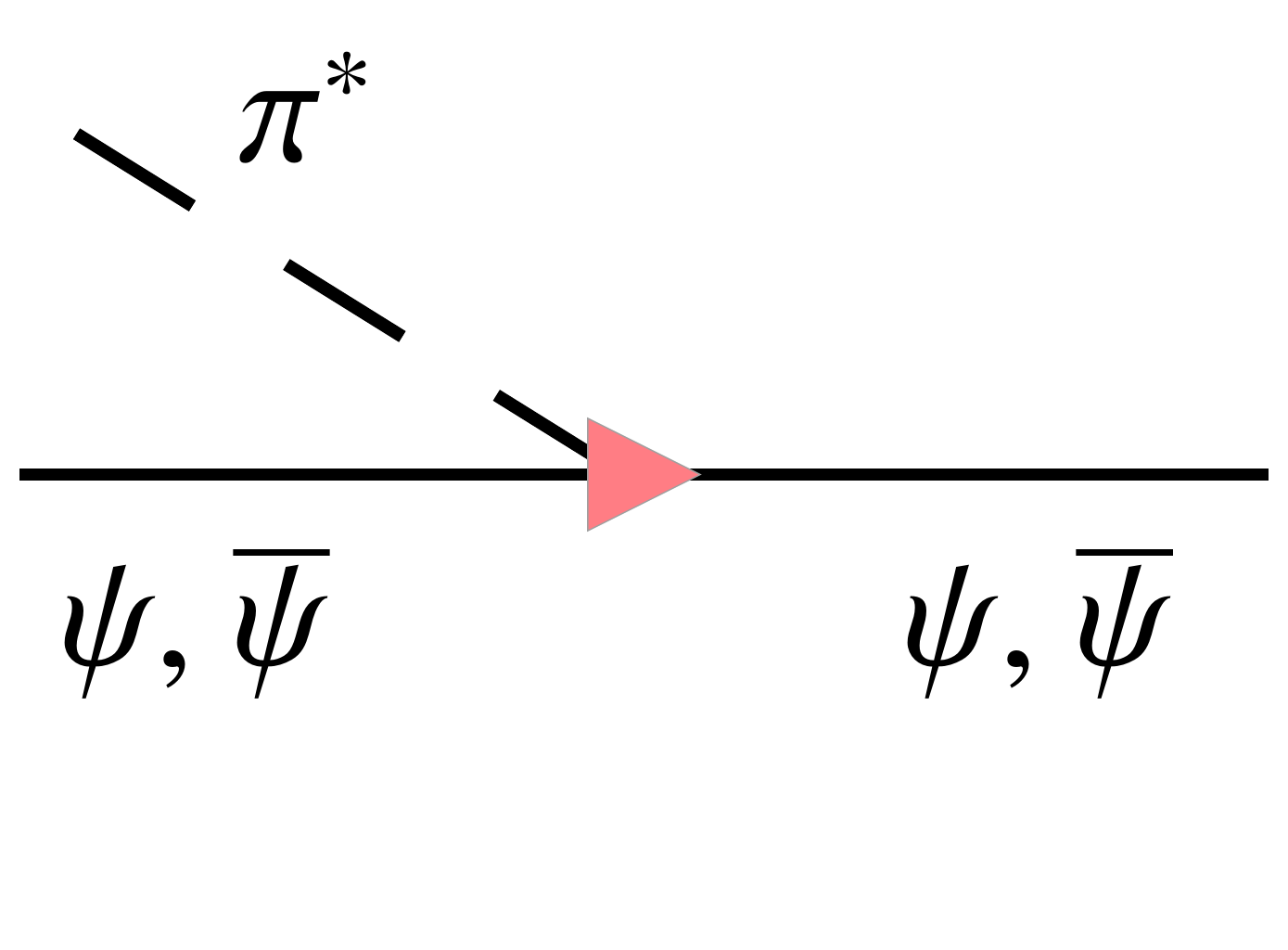}
}
	\caption{Graphical representation of the possible time-like, $p^2=\omega^2-\vec{p}^{\,2}>0$, and space-like, $p^2<0$, processes for an off-shell sigma, $\sigma^*$, or pion, $\pi^*$. First column: time-like decay channels for a sigma meson. Second column: time-like decay channels for a pion. Third column: absorption or emission of a space-like sigma excitation. Fourth column: absorption or emission of a space-like pion excitation. The Figure is taken from \cite{Tripolt2014a}.}
	\label{fig:processes} 
\end{figure}

In this section we present results for the mesonic spectral functions of the QM model. The structure of these spectral functions is determined by the different physical processes that can occur within our truncation, cf.~Fig.~\ref{fig:processes}. We divide these processes into two groups: time-like processes where the energy $\omega$ of the external particle is larger than its spatial momentum $|\vec{p}|$, and space-like processes with $\omega<|\vec{p}|$. The kinematic thresholds of these processes are determined by energy- and momentum-conservation, see also \cite{Tripolt2014a, Tripolt2015}.

\begin{figure}
	\centerline{
		{\includegraphics[width=0.5\textwidth]{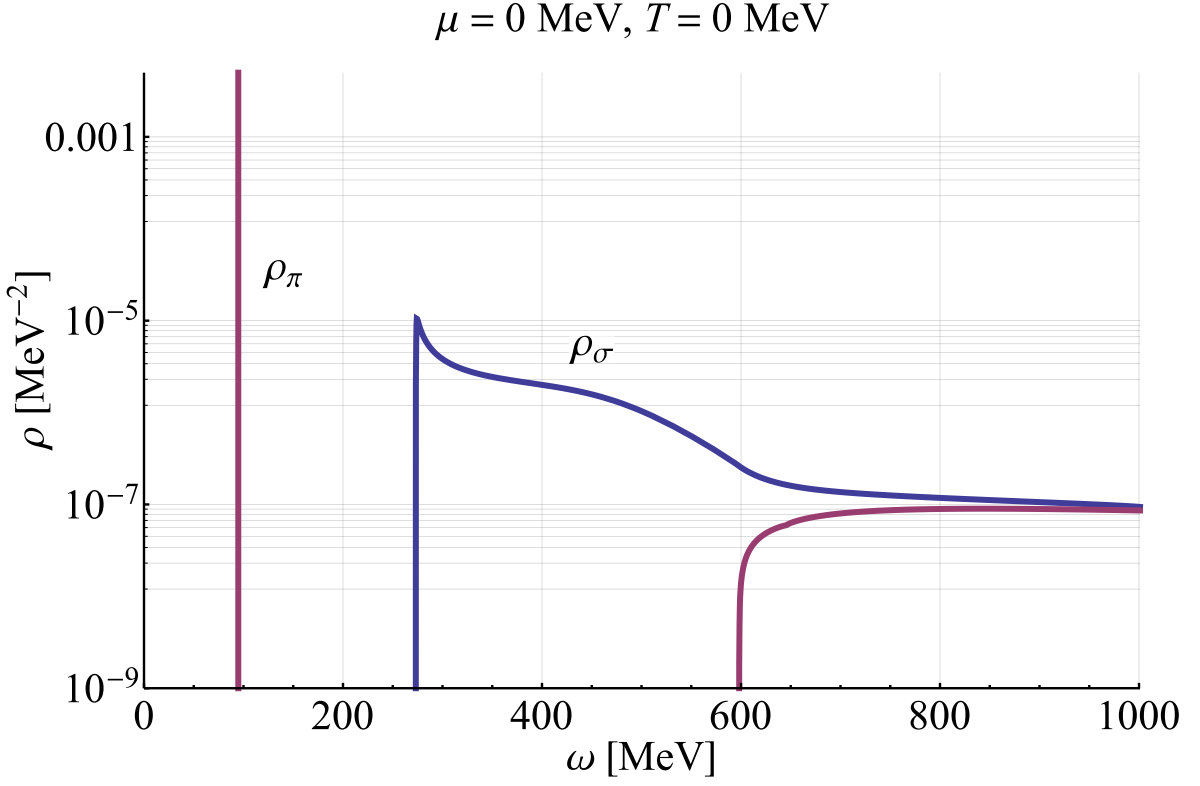}}
		{\includegraphics[width=0.5\textwidth]{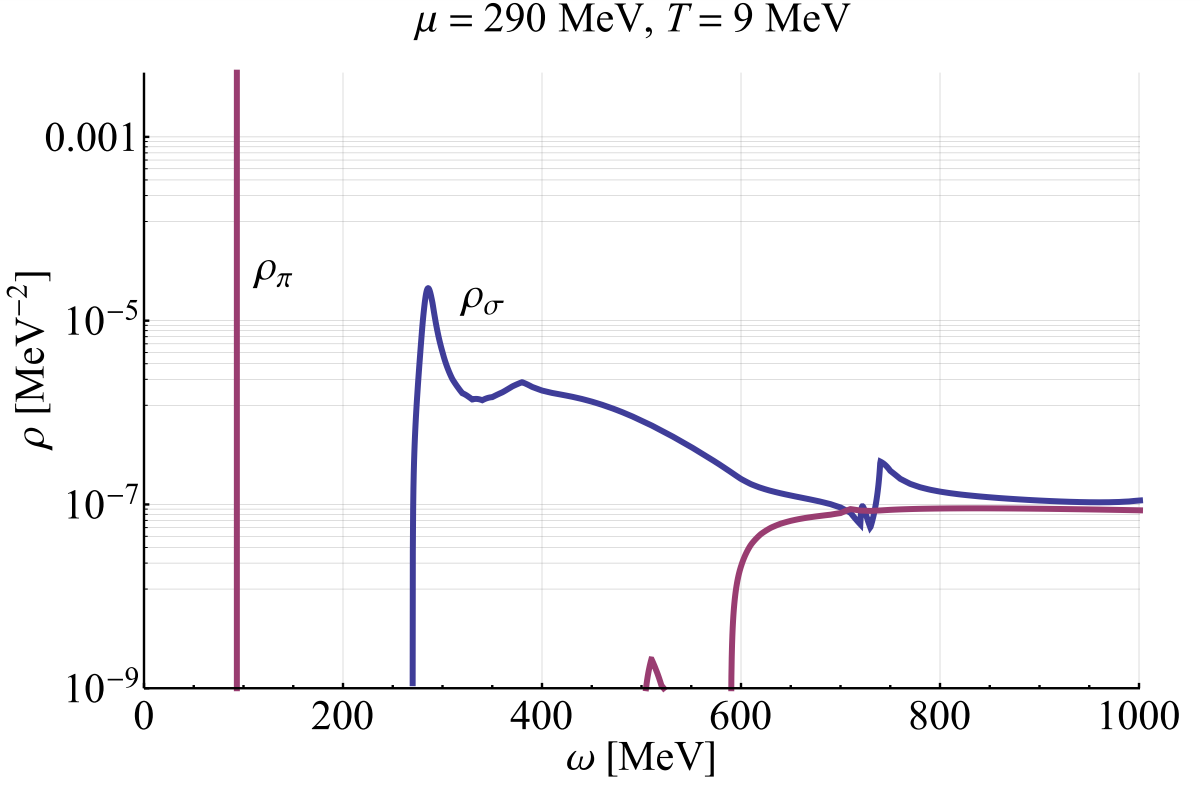}}
	}
	\vspace{2mm}
	\centerline{
		{\includegraphics[width=0.5\textwidth]{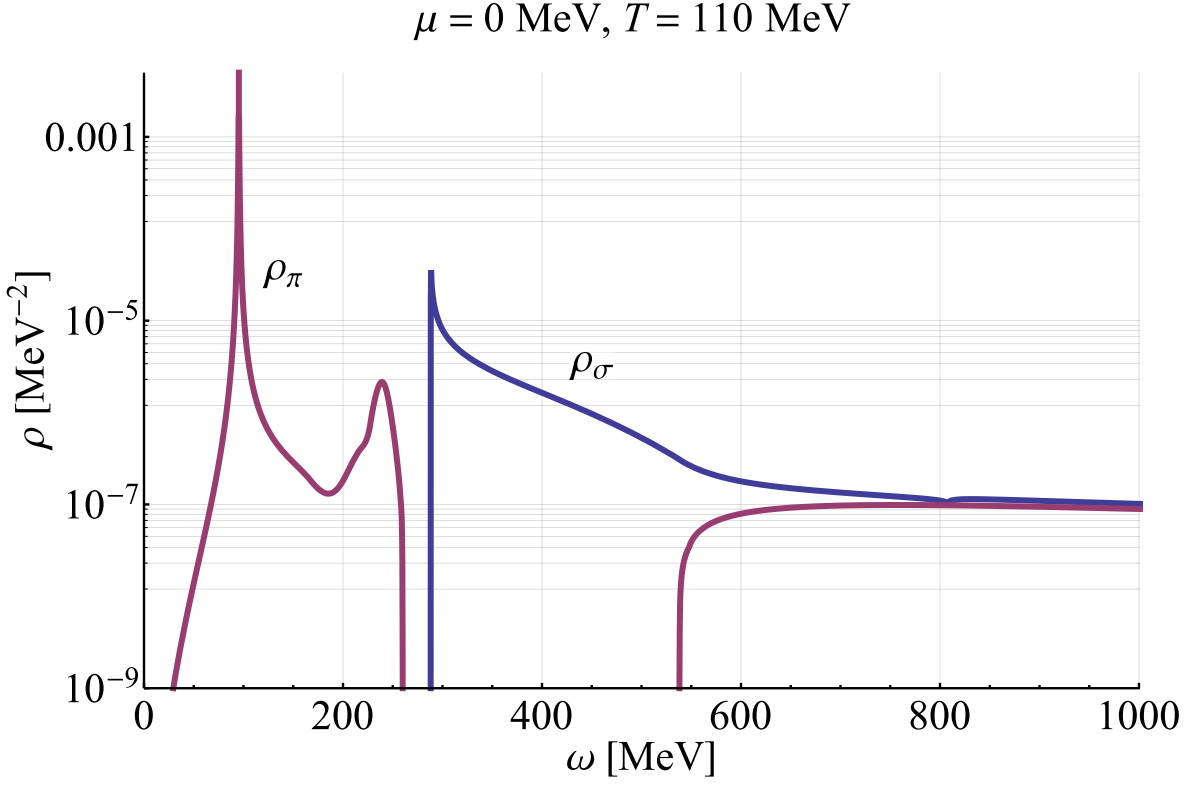}}
		{\includegraphics[width=0.5\textwidth]{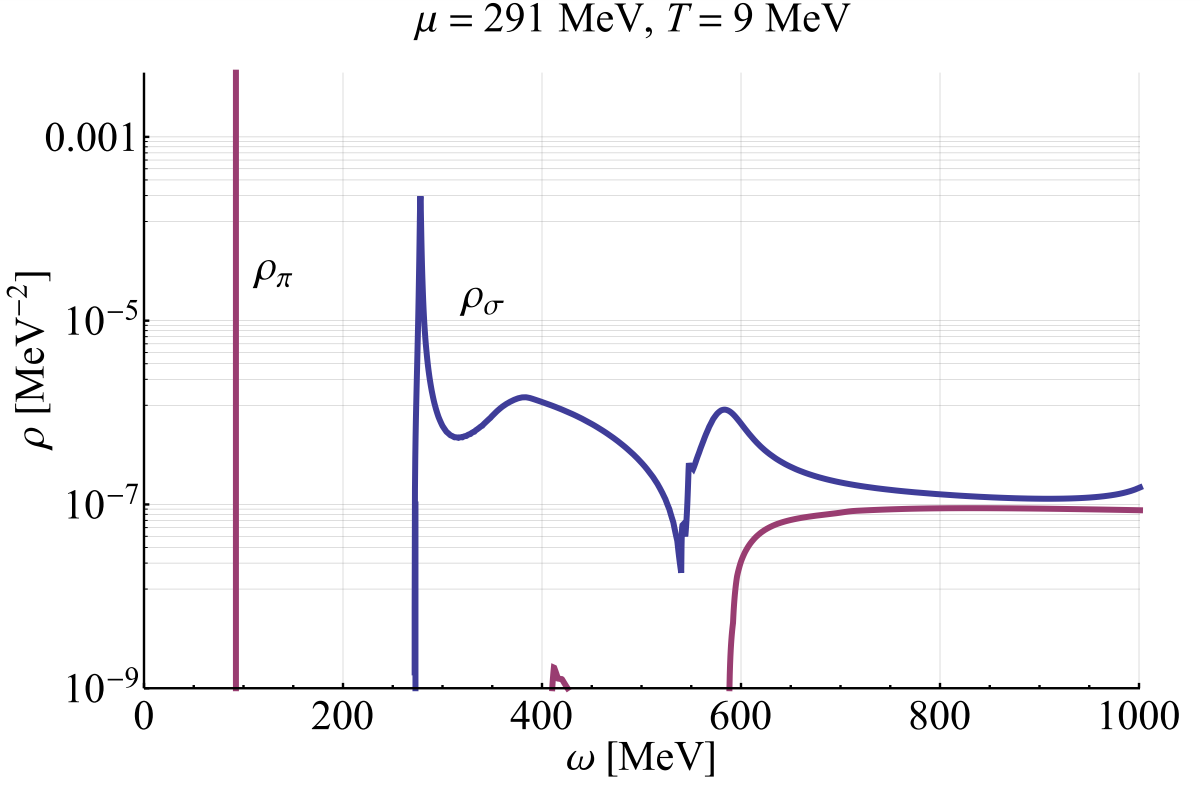}}
	}
	\vspace{2mm}
	\centerline{
		{\includegraphics[width=0.5\textwidth]{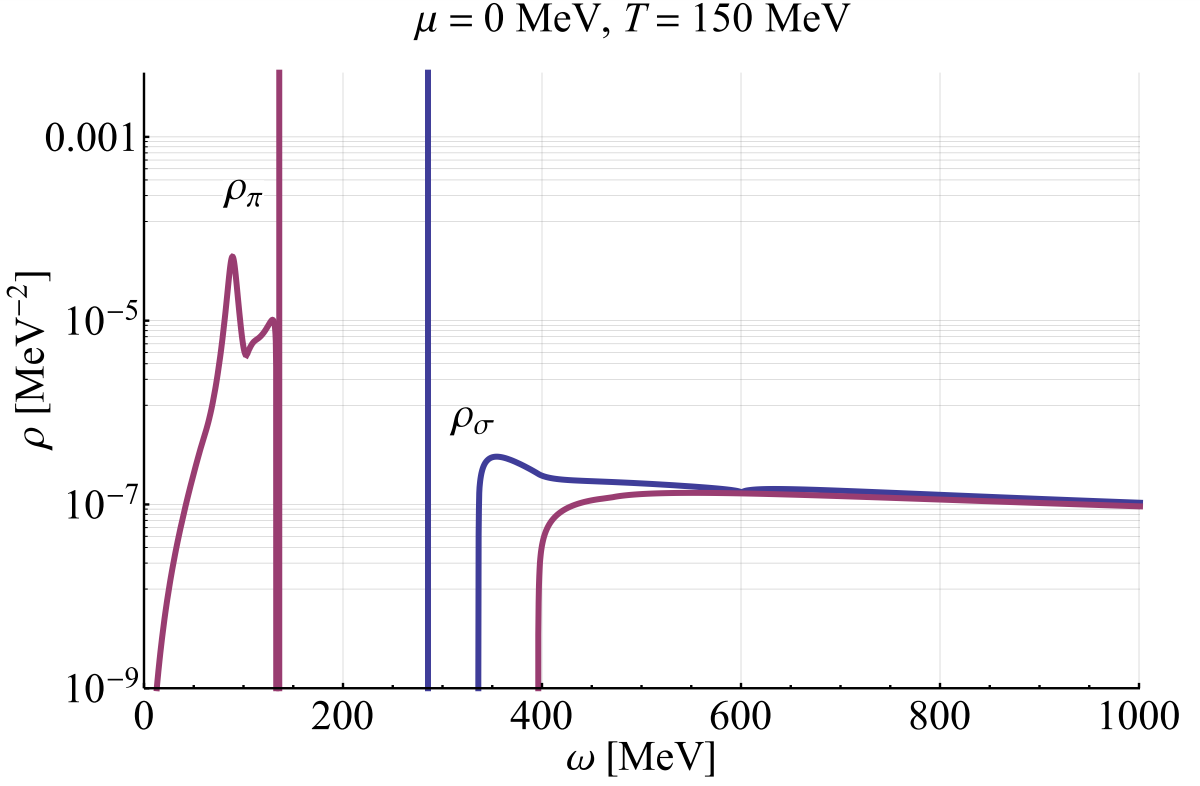}}
		{\includegraphics[width=0.5\textwidth]{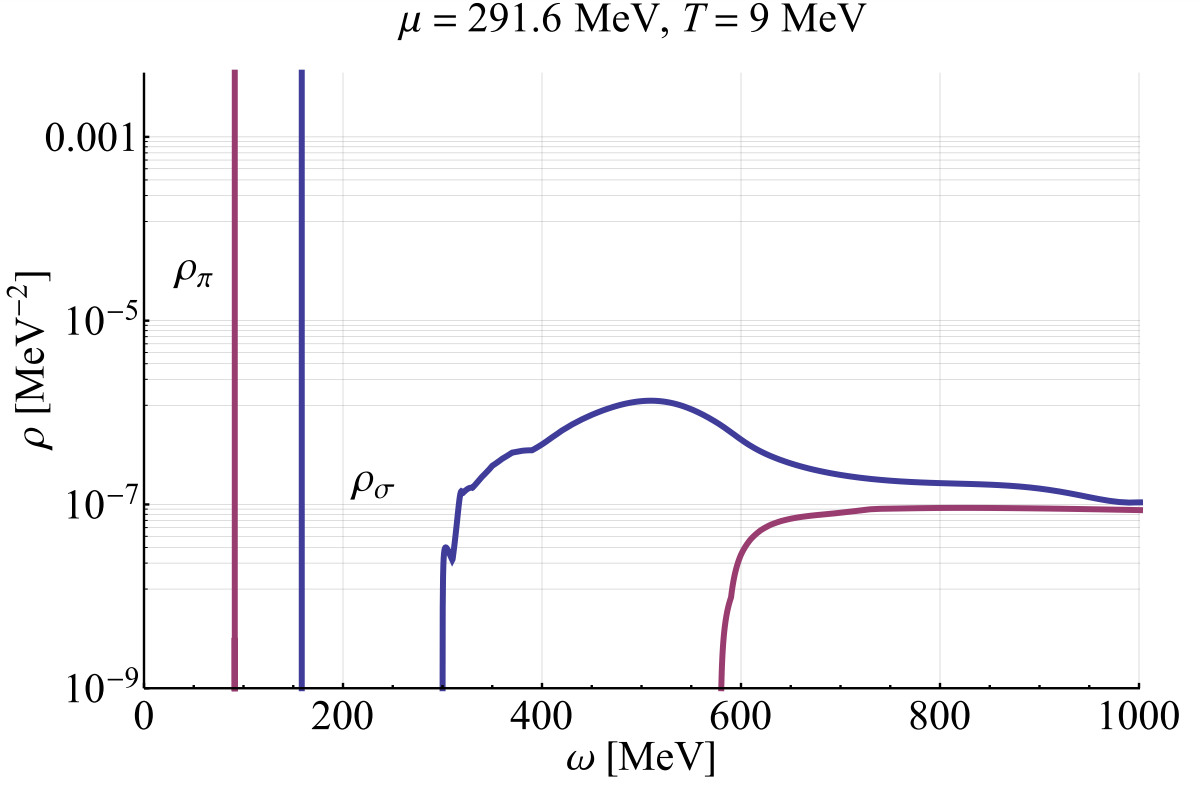}}
	}
	\vspace{2mm}
	\centerline{
		{\includegraphics[width=0.5\textwidth]{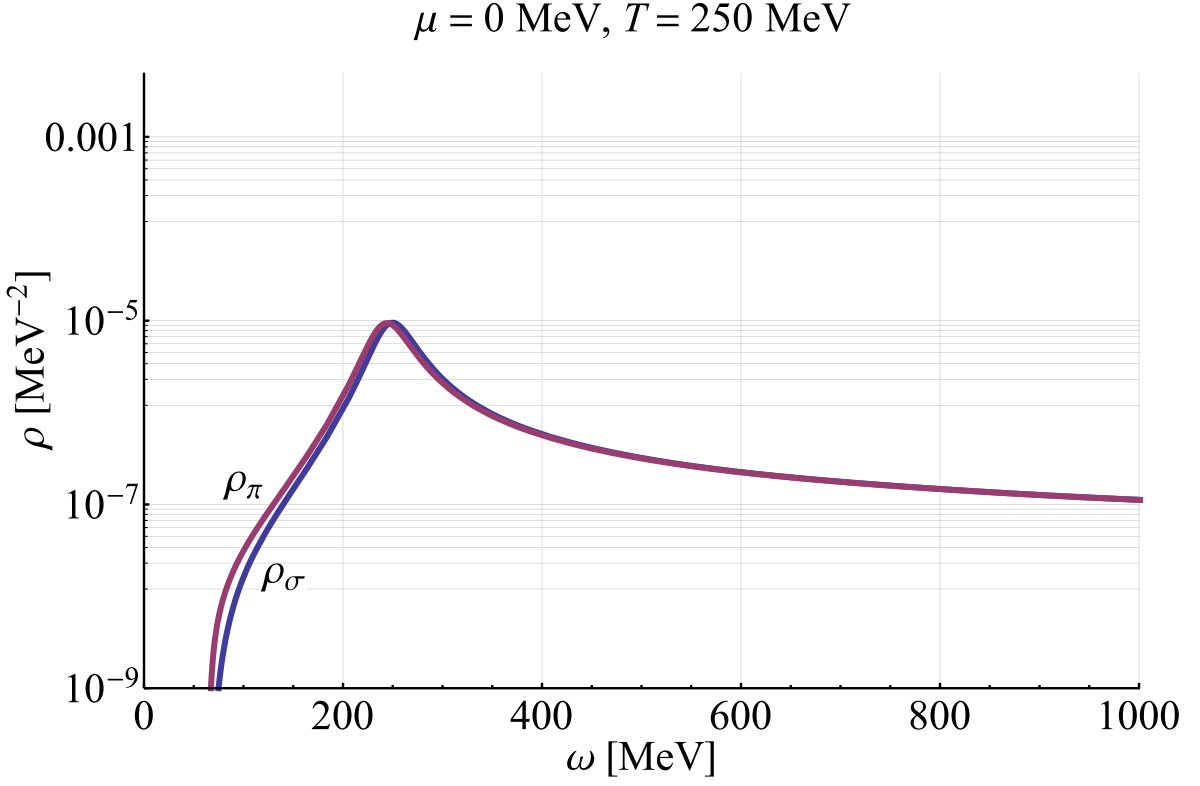}}
		{\includegraphics[width=0.5\textwidth]{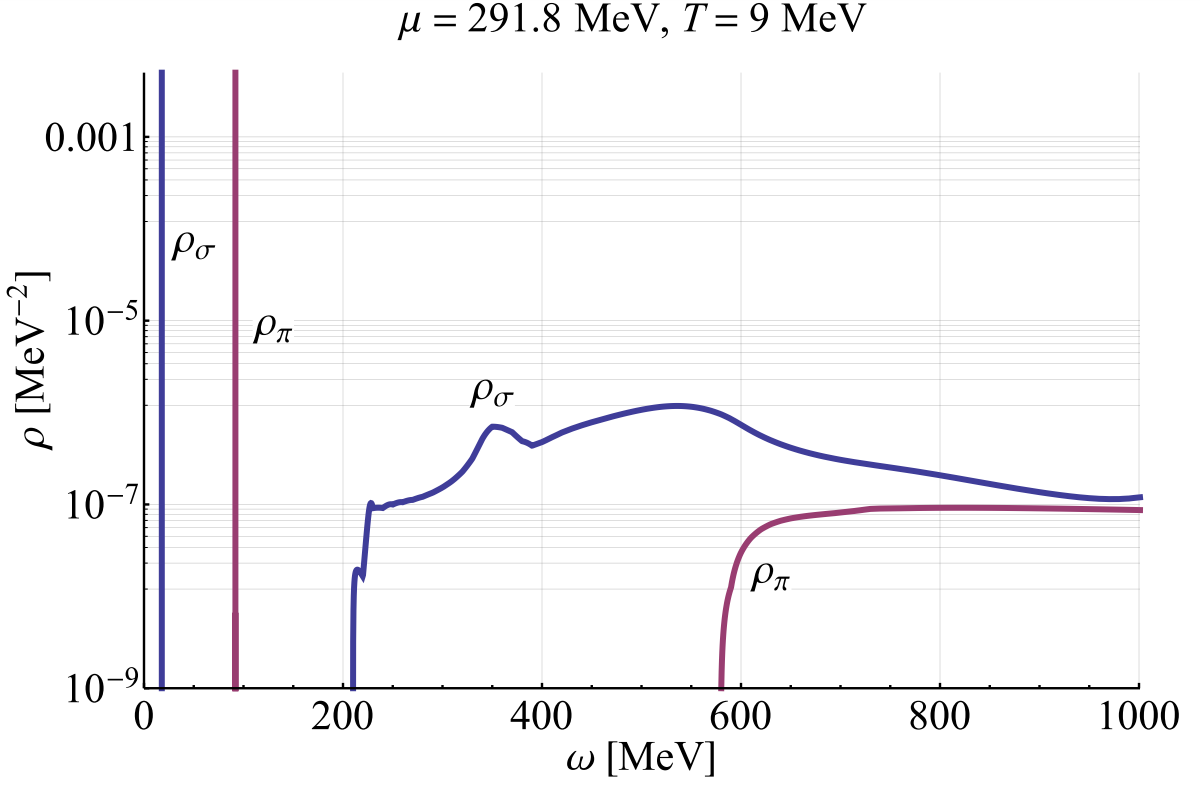}}
	}
	\caption{The in-medium sigma and pion spectral functions, $\rho_\sigma$ and $\rho_\pi$, are shown vs. energy $\omega$ at $|\vec{p}|=0$ for different temperatures and chemical potentials. Left panel: $\mu=0$ and $T$ increases from top to bottom. Right panel: $T=9$~MeV and $\mu$ increases from top to bottom, thus approaching the critical endpoint.}
	\label{fig:spectral_mu_T}
\end{figure}

In Fig.~\ref{fig:spectral_mu_T} we show the spectral functions for pions and sigma mesons which are at rest relative to the heat bath, {\it i.e.} $|\vec{p}|=0$, for different combinations of temperature and chemical potential. The left panel shows the evolution along the temperature axis in the phase diagram and the right panel the evolution along the $T=9$~MeV-line towards the CEP. 

\enlargethispage*{3mm}
As shown in Sect.~\ref{sec:vacuum} the stable vacuum pion is represented by a Dirac $\delta$-function whose location defines the pole mass. The sigma meson in the vacuum is unstable since it can decay into two pions, $\sigma^*\to \pi+\pi$. At higher energies, other processes such as the decay into a quark-antiquark pair give rise to modifications of the spectral functions. At $T=110$~MeV, the pion has become unstable since it can capture a pion from the heat bath and turn into a sigma meson, $\pi^*+\pi\to\sigma$. At $T=150$~MeV, i.e. near the crossover transition, also the sigma meson becomes stable since it can no longer decay into two pions. At higher temperatures beyond the crossover, the mesons become unstable again, since they can decay into light quark-antiquark pairs, and the spectral functions eventually become identical. It should be noted that complete degeneracy is only reached at temperatures significantly higher than the (pseudo) critical temperature derived from the evolution of the chiral condensate. This implies that chiral correlations persist well into the restored phase, as can also be inferred from the corresponding evolution of the curvature masses in Fig.~\ref{fig:masses}.

When looking at the dependence of the spectral functions on the chemical potential at $T=9$~MeV, we observe that they remain unchanged over a wide range of $\mu$, as expected from the Silver Blaze property \cite{Cohen2003}. Near the CEP, however, the sigma spectral function undergoes significant changes, while the pion spectral function remains almost unchanged. In particular, we observe that the sigma meson becomes stable and almost massless near the CEP as it should be.

\begin{figure}
\centerline{
{\includegraphics[width=0.55\textwidth]{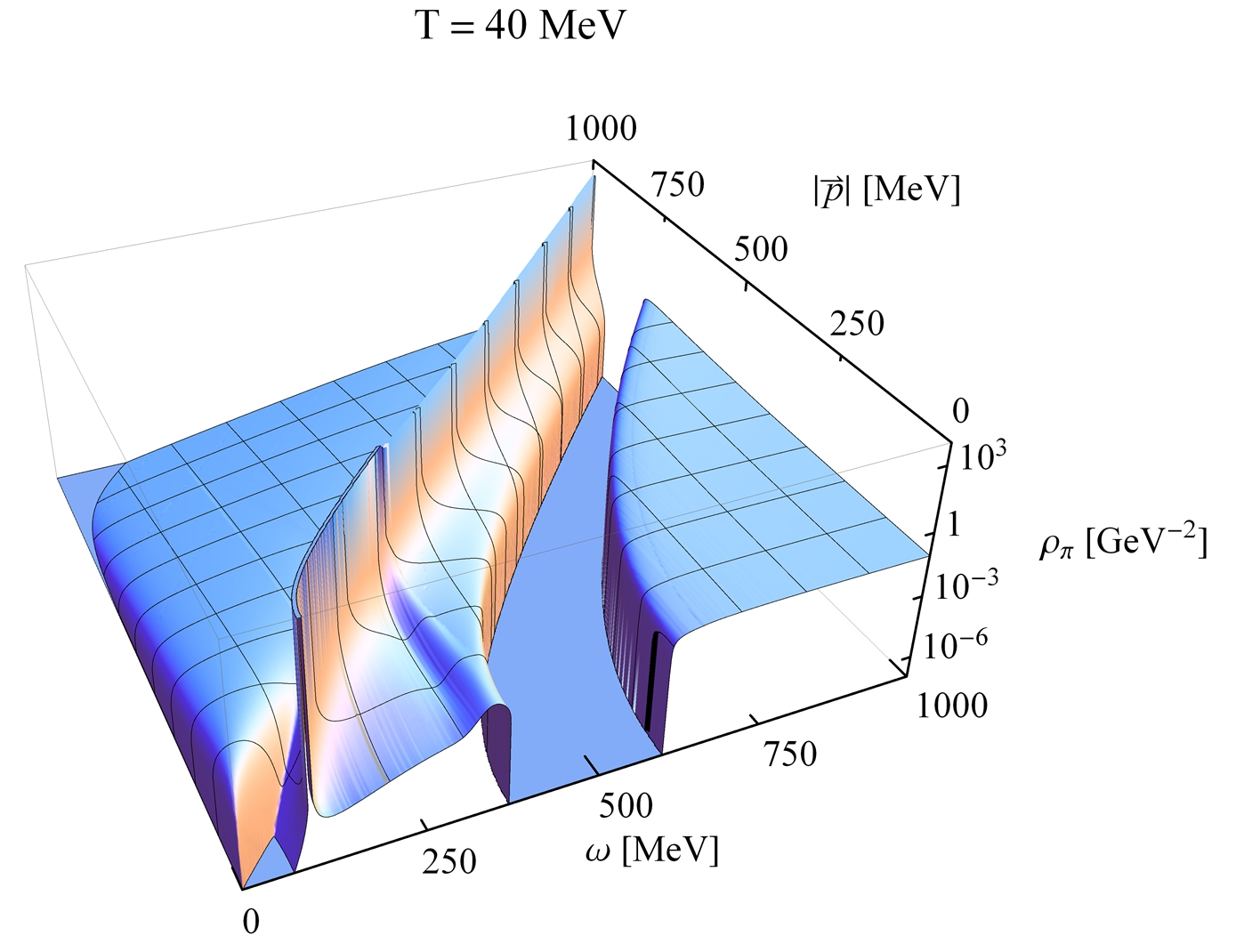}}
\hspace*{-0.03\textwidth}
{\includegraphics[width=0.55\textwidth]{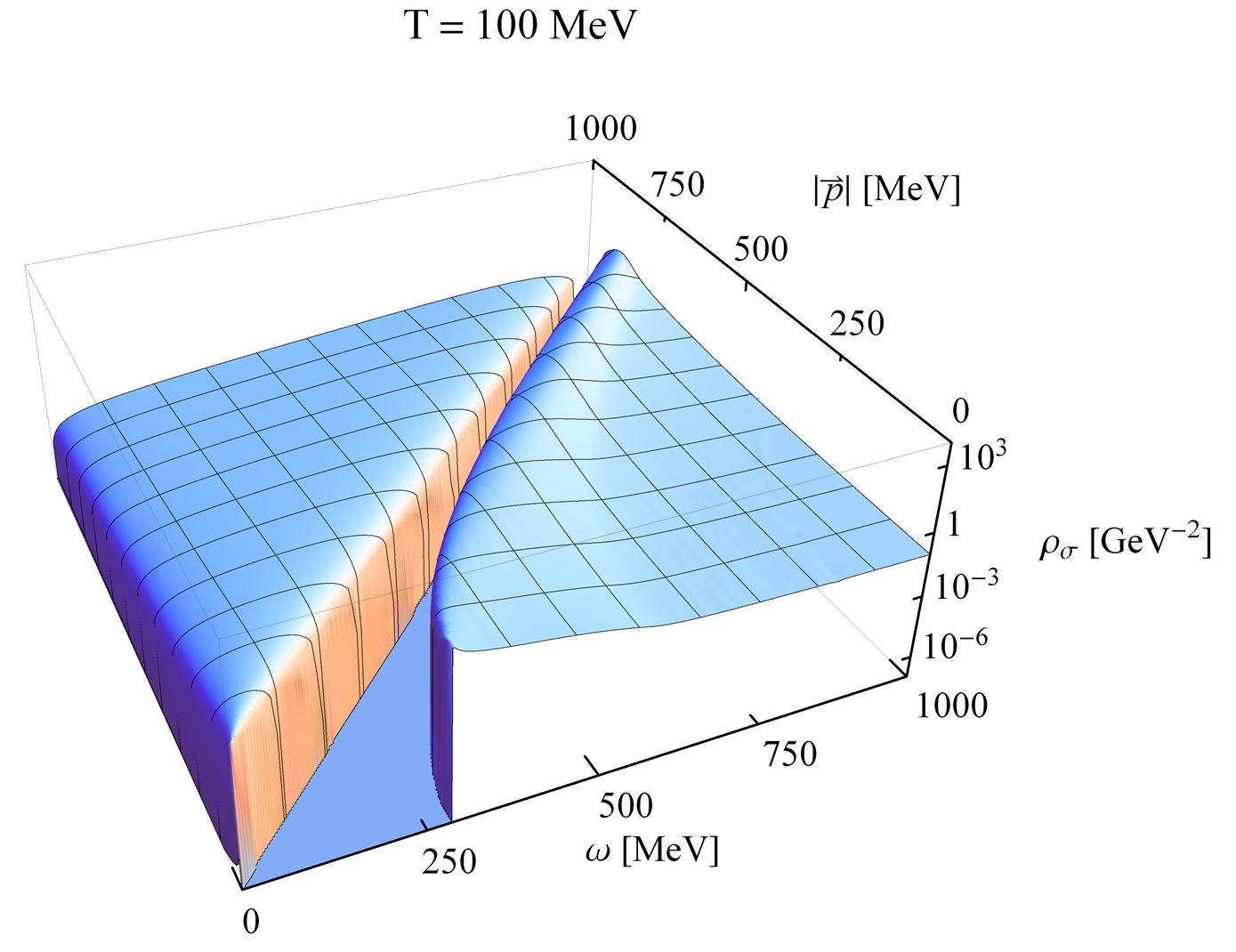}}
}
\caption{Left: The pion spectral function is shown vs.~energy $\omega$ and spatial momentum $|\vec{p}|$ at $T=40$~MeV and $\mu=0$. Right: The sigma spectral function is shown vs.~$\omega$ and $|\vec{p}|$ at $T=100$~MeV and $\mu=0$. }
\label{fig:spectral_3D}
\end{figure}

We finally turn to the full energy- and momentum-dependence of the spectral functions which contains information on all physical processes that can happen in the medium. They are also the essential input for computing the transport properties of the matter such as the shear viscosity. In Fig.~\ref{fig:spectral_3D}, the energy- and momentum-dependence of the pion and sigma spectral function is shown for finite temperatures and $\mu=0$. The space-like region ($|\vec{p}|>\omega$) is non-vanishing at finite temperature due to the various space-like processes, cf.~Fig.~\ref{fig:processes}, while the decay thresholds and the pion-peak within the time-like region ($|\vec{p}|<\omega$) are Lorentz-boosted to higher energies as the spatial momentum increases. 

\clearpage

\section{Shear viscosity and $\eta/s$}
\label{sec_eta}

As mentioned above, the spectral functions in Minkowski space-time contain the information of all physical scattering processes and they are thus suitable to determine the transport properties of the matter.  When evaluated in the FRG it is then possible to treat the influence of phase transitions reliably.

\begin{figure}
\centerline{
{\includegraphics[width=0.5\textwidth]{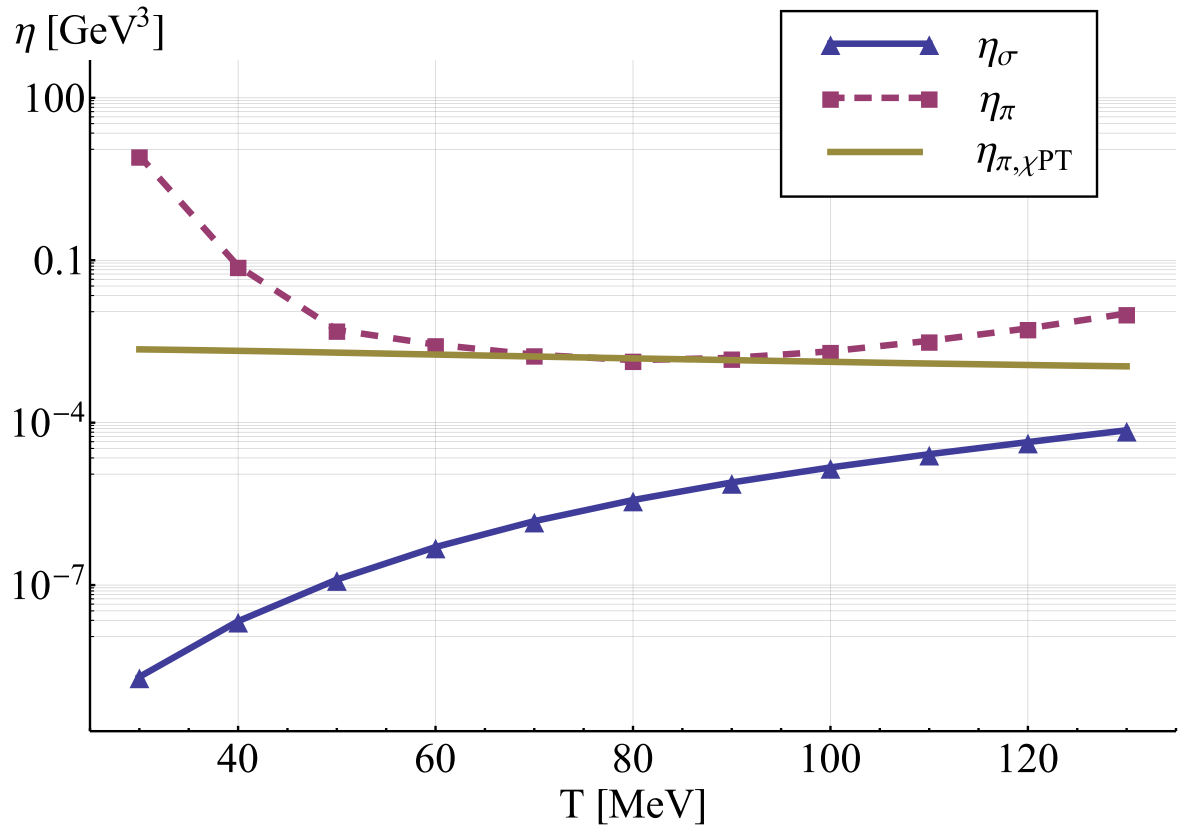}}
{\includegraphics[width=0.5\textwidth]{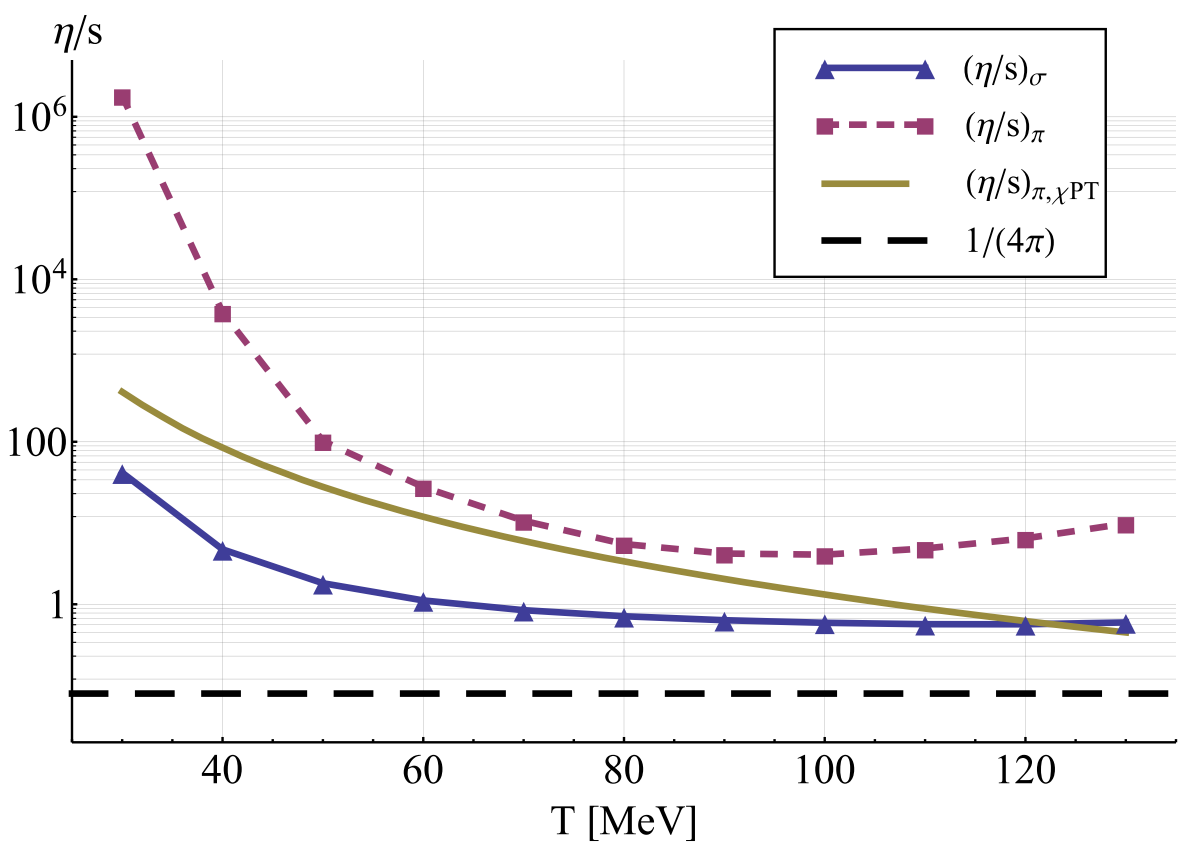}}
}
\caption{The shear viscosity (left) and the shear viscosity over entropy density ratio (right) of the sigma mesons and the pions are shown vs.~temperature at $\mu=0$ in comparison to results based on chiral perturbation theory ($\chi$PT) \cite{Lang2012}.}
\label{fig:eta}
\end{figure}

The suitable framework is the Green-Kubo formalism~\cite{Green1954, Kubo1957} of linear response theory. It is more general than kinetic theory since nowhere the quasiparticle picture has to be invoked. For the shear viscosity $\eta$, which is of great interest for the space-time evolution of relativistic heavy-ion collisions, one has
 
\begin{equation}
\eta=\frac{1}{24}\lim_{q_0 \rightarrow 0}\lim_{|\vec q| \rightarrow 0} \frac{1}{q_0}
\int d^4x \:\:e^{i q x} \left\langle\left[ T_{ij}(x), T^{ij}(0)\right]\right\rangle,
\end{equation}

\noindent
where $T_{ij}$ are the spatial components of the (local) energy-momentum tensor. Inserting the energy-momentum tensor of the quark-meson model, which can be straightforwardly evaluated, gives
\begin{equation}
\eta_{\sigma,\pi}\propto
\int\frac{d\omega}{2\pi}\int\frac{d^3p}{(2\pi)^3}\:\:
p_x^2\:p_y^2
\: n_B^\prime(\omega)
\: \rho_{\sigma,\pi}^2(\omega,\vec p),
\end{equation}
as the main contribution to the mesonic shear viscosities, where $n_B^\prime$ denotes the energy derivative of the Bose occupation factor. In Fig.~\ref{fig:eta} we show results for the shear viscosity as well as for $\eta/s$ of the mesons in comparison to a result based on chiral perturbation theory ($\chi$PT), \cite{Lang2012}. We find good agreement for $\eta$ and $\eta/s$ of the pions at intermediate temperatures, while towards lower temperatures, i.e.~$T\lesssim 50$~MeV, and higher temperatures, i.e.~$T\gtrsim 100$~MeV, our results for $\eta$ and $\eta/s$ are larger than expected from $\chi$PT. These differences partly stem from the fact that two-pion scattering processes are not yet included in our truncation. The decrease of $\eta_\sigma$ towards lower temperatures is simply a temperature effect, which, in the case of the pions, is overcompensated by a sharpening of the pion peak, cf.~Fig.~\ref{fig:spectral_3D}.

\section{Summary and outlook}
\label{sec:summary}

Spectral functions of physical systems in equilibrium contain the information on the measurable elementary excitations and, taking suitable limits, determine the transport properties. A well-known self-consistent method for evaluating spectral functions is the RPA, first introduced in the electron gas. In nuclear physics, Gerry has applied this method extensively for collective modes like the giant dipole resonance. However, the RPA is based on a mean-field picture, it has difficulties in properly dealing with phase transitions. Since the QCD phase diagram may contain such transitions as chiral symmetry gets restored with increasing temperature and density, it is therefore imperative to link the spectral properties of QCD matter to the modifications of the chiral order parameter in a framework that goes beyond the mean-field approximation. Such a method is the FRG in which the equilibrium thermodynamics is obtain from suitably truncated momentum-flow equations of the Grand Potential in Euclidean space-time.      

To compute spectral functions in this framework, flow equations for the 2-point functions have to be derived with suitable truncations for the pertinent vertex functions. These are then analytically continued to Minkowski space-time. We have recently proposed a truncation scheme that is thermodynamically consistent and symmetry conserving. The latter is important to properly account for the Goldstone nature of spontaneously broken symmetries that is chiral symmetry in QCD. Within the Quark-Meson Model as an effective theory for QCD, we have applied our truncation scheme to the scalar parity partners, the pion and the sigma meson. Starting from the unbroken phase at the ultraviolet cutoff, for which both mesons have large and practically degenerate masses, the momentum flow generates spontaneous chiral symmetry breaking and the pion and sigma spectral functions
split. In the vacuum the decay of the sigma mesons into two pions as the dominant channel is obtained naturally. As a consequence of chiral symmetry restoration in the medium both the pion and sigma meson spectral functions are modified in a complicated fashion and eventually become identical when the symmetry is fully restored. 

Let us come back to Gerry. He and his collaborators became interested early on in the question of how spontaneous chiral symmetry breaking and its restoration with temperature and density influences the spectral properties of vector mesons. The $\rho$-meson was of special interest since it strongly couples to the $e^+e^-$ channel and hence its in-medium modification should be observable in the di-lepton rates of relativistic heavy-ion collisions.  This is a long story (see \cite{Rapp:2009yu} for a recent review) but the question persists to which extent the observed effects are indicators for the restoration of chiral symmetry. From the discussions above it is obvious that in a consistent chiral theory the parity partner of the $\rho$ meson has to be treated on equal footing \cite{Hohler:2015iba}. Chiral symmetry restoration then manifests itself in the complete degeneracy of the spectral functions of the parity partners. It should be clear that the FRG approach outlined above is suitable to address this question. Based on a gauged Quark-Meson Model \cite{Urban:2001ru} we have started to tackle this question and first promising results have been obtained \cite{C.Jung}. We are sure that Gerry would have appreciated them.

\section*{Acknowledgments}
We dedicate this article to Gerry Brown, an inspiring `father figure' for generations of theoretical nuclear physicists and a great human being. 
This work was supported by the BMBF, project number 05P12RDGHD, and the Helmholtz International Center for FAIR within the LOEWE initiative of the state of Hesse. R.-A.~T. was furthermore supported by the Helmholtz Research School for Quark Matter Studies, H-QM.

\bibliographystyle{ws-rv-van}
\bibliography{Gerry_Brown_World_Scientific_JW.bbl}

\end{document}